\documentclass[useAMS,usenatbib,letterpaper]{mn2e}
\usepackage{graphicx}
\usepackage{epstopdf}
\usepackage{epsfig}
\usepackage{subfigure}
\usepackage{hyperref}
\usepackage{bm}
\usepackage{natbib}
\usepackage{color}
\usepackage[mediumspace,Gray,squaren]{SIunits}
\usepackage{multirow}
\usepackage{comment}
\usepackage{amsmath}
\usepackage[pass]{geometry}
\usepackage{aas_macros}
\usepackage[T1]{fontenc}
\usepackage{aecompl}
\title[Millimeter Emission and SZ Effect Associated with Radio Sources]{A Measurement of the Millimeter Emission and the Sunyaev-Zel'dovich Effect Associated with Low-Frequency Radio Sources}
\author[Gralla et al.]{
Megan B. Gralla,$^{1}$
Devin Crichton,$^{1}$
Tobias A. Marriage,$^{1}$
Wenli Mo,$^{1,2}$
\newauthor
Paula Aguirre,$^{3}$
Graeme E. Addison,$^{4}$
V. Asboth,$^{4}$
Nick Battaglia,$^{5}$
James Bock,$^{6,7}$
\newauthor
J. Richard Bond,$^{8}$
Mark J. Devlin,$^{9}$
Rolando D\"unner,$^{3}$
Amir Hajian,$^{8}$
Mark Halpern,$^{4}$
\newauthor
Matt Hilton,$^{10,11}$
Adam D. Hincks,$^{4}$
Ren\'ee A. Hlozek,$^{12}$
Kevin M. Huffenberger,$^{13}$
\newauthor
John P. Hughes,$^{14}$
R.\,J.~Ivison,$^{15,16}$
Arthur Kosowsky,$^{17}$
Yen-Ting Lin,$^{18}$
\newauthor
Danica Marsden,$^{19}$
Felipe Menanteau,$^{20}$
Kavilan Moodley,$^{10}$
Gustavo Morales,$^{3}$
\newauthor
Michael D. Niemack,$^{21}$
Seb Oliver,$^{22}$
Lyman A. Page,$^{23}$
Bruce Partridge,$^{24}$
\newauthor
Erik D. Reese,$^{9,25}$
Felipe Rojas,$^{3}$
Neelima Sehgal,$^{26}$
Jon Sievers,$^{10,23}$
Crist\'obal Sif\'on,$^{27}$
\newauthor
David N. Spergel,$^{12}$
Suzanne T. Staggs,$^{23}$
Eric R. Switzer,$^{8,28}$
Marco P. Viero,$^{6}$
\newauthor
Edward J. Wollack,$^{28}$
Michael B. Zemcov$^{6,7}$\vspace{0.4cm}\\
\parbox{\textwidth}{
$^{1}$Dept. of Physics and Astronomy, Johns Hopkins University, 3400 N. Charles St., Baltimore, MD 21218\\
$^{2}$Dept. of Astronomy, University of Florida, Gainesville, FL 32611\\
$^{3}$Instituto de Astrof\'isica, Facultad de F\'isica, Pontificia Universidad Cat\'olica de Chile, Casilla 306, Santiago 22, Chile\\
$^{4}$Department of Physics and Astronomy, University of British Columbia, 6224 Agricultural Road, Vancouver, BC V6T 1Z1, Canada\\
$^{5}$McWilliams Center for Cosmology, Wean Hall, Carnegie Mellon University, 5000 Forbes Ave., Pittsburgh PA 15213, USA\\
$^{6}$California Institute of Technology, 1200 East California Boulevard., Pasadena, CA 91125, USA\\
$^{7}$Jet Propulsion Laboratory, 4800 Oak Grove Drive, Pasadena, CA 91109, USA\\
$^{8}$Canadian Institute for Theoretical Astrophysics, University of Toronto, Toronto, ON, M5S 3H8, Canada\\
$^{9}$Department of Physics and Astronomy, University of Pennsylvania, 209 South 33rd Street, Philadelphia, PA 19104, USA\\
$^{10}$Astrophysics and Cosmology Research Unit, School of Mathematics, Statistics \& Computer Science, University of KwaZulu-Natal, Durban, 4041, South Africa\\
$^{11}$Centre for Astronomy \& Particle Theory, School of Physics \& Astronomy, University of Nottingham, Nottingham, NG7 2RD, U.K\\
$^{12}$Department of Astrophysical Sciences, Peyton Hall, Princeton University, Princeton, NJ 08544, USA\\
$^{13}$Department of Physics, Florida State University, Tallahassee, Florida 32306\\
$^{14}$Department of Physics and Astronomy, Rutgers, The State University of New Jersey, Piscataway, NJ 08854-8019, USA\\
$^{15}$UK Astronomy Technology Centre, Science and Technology Facilities Council, University of Edinburgh\\
$^{16}$Institute for Astronomy, University of Edinburgh, Royal Observatory, Blackford Hill, Edinburgh EH9 3HJ, UK\\
$^{17}$Department of Physics and Astronomy, University of Pittsburgh, Pittsburgh, PA 15260, USA\\
$^{18}$Institute of Astronomy and Astrophysics, Academia Sinica, Taipei, Taiwan\\
$^{19}$Department of Physics, University of California Santa Barbara, CA 93106, USA\\
$^{20}$National Center for Supercomputing Applications, University of Illinois at Urbana-Champaign, 1205 W. Clark St, Urbana IL, 61801\\
$^{21}$Department of Physics, Cornell University, Ithaca, NY 14853\\
$^{22}$Astronomy Centre, Department of Physics \& Astronomy, University of Sussex, Brighton BN1 9QH, UK\\
$^{23}$Joseph Henry Laboratories of Physics, Jadwin Hall, Princeton University, Princeton, NJ, USA 08544\\
$^{24}$Department of Physics and Astronomy, Haverford College, Haverford, PA 19041, USA\\
$^{25}$Department of Physics, Astronomy, and Engineering, Moorpark College, 7075 Campus Rd., Moorpark, CA 93021\\
$^{26}$Department of Physics and Astronomy, Stony Brook University, Stony Brook, NY 11794-3800, USA\\
$^{27}$Leiden Observatory, Leiden University, P.O. Box 9513, NL-2300 RA Leiden, The Netherlands\\
$^{28}$NASA/Goddard Space Flight Center, Greenbelt, MD, 20771, USA
}}

\date{Accepted 5 August 2014}

\pagerange{\pageref{firstpage}--\pageref{lastpage}} \pubyear{2013}

\def\LaTeX{L\kern-.36em\raise.3ex\hbox{a}\kern-.15em
    T\kern-.1667em\lower.7ex\hbox{E}\kern-.125emX}

\begin{document}
\newcommand{\AROnePixwin}{1.06}
\newcommand{\ARTwoPixwin}{1.10}
\newcommand{\ARThreePixwin}{1.14}

\newcommand{\TotalInRegionSources}{9,436}
\newcommand{\TotalInFluxRangeSources}{4,563}
\newcommand{\ACTExcluded}{219}
\newcommand{\ACTExcludedPercent}{4.8\%}
\newcommand{\TotalStackedSources}{4,344}
\newcommand{\SourcesWithIR}{2,123}
\newcommand{\NullChiSqAROne}{5.7}
\newcommand{\NullDOFAROne}{7}
\newcommand{\NullPTEAROne}{0.57}
\newcommand{\NullChiSqARTwo}{11.3}
\newcommand{\NullDOFARTwo}{7}
\newcommand{\NullPTEARTwo}{0.13}
\newcommand{\NullChiSqARThree}{3.4}
\newcommand{\NullDOFARThree}{7}
\newcommand{\NullPTEARThree}{0.85}
\newcommand{\NBins}{7}
\newcommand{\ParASZWithSZSignif}{5}
\newcommand{\FinalModelDir}{plots}

\newcommand{\LowBinCount}{1,767}
\newcommand{\HiBinCount}{61}
\newcommand{\LowBinFlux}{6.4}
\newcommand{\HiBinFlux}{149.1}

\newcommand{\SPivmJy}{149.1}
\newcommand{\ModelChiSqWithSZ}{31.1}
\newcommand{\ModelDOFWithSZ}{27}
\newcommand{\ModelPTEWithSZ}{0.27}
\newcommand{\ParASZWithSZ}{$0.306 \pm 0.052$}
\newcommand{\ParAlphaWithSZ}{$0.95 \pm 0.01$}
\newcommand{\ParBetaWithSZ}{$0.31 \pm 0.02$}
\newcommand{\ParLBolWithSZ}{$9.96 \pm 0.08$}

\newcommand{\ModelChiSqWithSZandSteep}{31.0}
\newcommand{\ModelDOFWithSZandSteep}{25}
\newcommand{\ModelPTEWithSZandSteep}{0.19}
\newcommand{\ParSWithSZandSteep}{$1.05 \pm 0.30$}
\newcommand{\ParASZWithSZandSteepSignif}{5}
\newcommand{\ParASZWithSZandSteep}{$0.304 \pm 0.057$}
\newcommand{\ParLBolWithSZandSteep}{$9.97 \pm 0.08$}
\newcommand{\ParAlphaWithSZandSteep}{$0.94 \pm 0.02$}
\newcommand{\ParBetaWithSZandSteep}{$0.31 \pm 0.02$}

\newcommand{\ModelChiSqWithoutSZandSteep}{55.3}
\newcommand{\ModelDOFWithoutSZandSteep}{26}
\newcommand{\ModelPTEWithoutSZandSteep}{$7 \times 10^{-4}$}
\newcommand{\ParSWithoutSZandSteep}{$1.29 \pm 0.22$}
\newcommand{\ParLBolWithoutSZandSteep}{$10.10 \pm 0.05$}

\newcommand{\GMBCGInReg}{1,903}
\newcommand{\GMBCGAssocSourcesPreThin}{405}
\newcommand{\GMBCGAssocSourcesPostThin}{192}
\newcommand{\GMBCGFracChangeAROne}{$< 5\%$}
    
\newcommand{\BestSFGCutNum}{211}
    
\newcommand{\Rfh}{0.25}
\newcommand{\Tfh}{0.64}
\newcommand{\AROneMeanCorr}{23\%}
\newcommand{\AROneMedCorr}{7\%}
\newcommand{\ARThreeMeanCorr}{37\%}
\newcommand{\ARThreeMedCorr}{10\%}
\newcommand{\AROneMeanCorrFac}{1.23}
\newcommand{\AROneMedCorrFac}{1.07}
\newcommand{\ARThreeMeanCorrFac}{1.37}
\newcommand{\ARThreeMedCorrFac}{1.10}
\newcommand{\AROneToARThreeMeanCorr}{12\%}
\newcommand{\YIntegratedAmpl}{$5.4 \pm 1.2^{stat} \pm 3^{sys} \times 10^{-8}$}
\newcommand{\YIntegratedAmplnosys}{$5.4 \pm 1.2^{stat} \times 10^{-8}$}
\newcommand{\YIntegratedAmplHo}{$4.5 \pm 1.0 \times 10^{-8}$}
\newcommand{\YIntegratedAmplPlanck}{$8.7 \pm 2.0 \times 10^{-8}$}
\newcommand{\YIntegratedAmplNeto}{$5.7 \pm 1.3 \times 10^{-8}$}

\newcommand{\RfhBest}{0.34}
\newcommand{\TfhBest}{1.24}
\newcommand{\AROneMeanCorrBest}{153\%}
\newcommand{\AROneMedCorrBest}{32\%}
\newcommand{\ARThreeMeanCorrBest}{271\%}
\newcommand{\ARThreeMedCorrBest}{52\%}
\newcommand{\AROneMeanCorrFacBest}{2.53}
\newcommand{\AROneMedCorrFacBest}{1.32}
\newcommand{\ARThreeMeanCorrFacBest}{3.71}
\newcommand{\ARThreeMedCorrFacBest}{1.52}
\newcommand{\AROneToARThreeMedCorrBest}{15\%}
\newcommand{\YIntBest}{$1.4 \pm 0.5^{stat} \pm 0.6 ^{sys} \times 10^{-7}$}
\newcommand{\YIntBestnosys}{$1.4 \pm 0.5^{stat}\times 10^{-7}$}
\newcommand{\YIntBestPlanck}{$2.0 \pm 0.7 \times 10^{-7}$ }
\newcommand{\YIntBestNeto}{$1.5 \pm 0.5 \times 10^{-7}$}

\newcommand{\KimballPercentThin}{35\%}
\newcommand{\BestPercentThin}{1.1\%}
    
\label{firstpage}
\maketitle
\clearpage
\begin{abstract}

We present a statistical analysis of the millimeter-wavelength properties of 1.4~GHz-selected sources and a detection of the Sunyaev-Zel'dovich effect associated with the halos that host them.  We stack data at 148, 218 and 277~GHz from the Atacama Cosmology Telescope at the positions of a large sample of radio AGN selected at 1.4~GHz.  The thermal Sunyaev-Zel'dovich (SZ) effect associated with the halos that host the AGN is detected at the \ParASZWithSZSignif$\sigma$ level through its spectral signature, representing a statistical detection of the SZ effect in some of the lowest mass halos (average $M_{200}\approx10^{13}$~M$_{\odot}h_{70}^{-1}$) studied to date. The relation between the SZ effect and mass (based on weak lensing measurements of radio galaxies) is consistent with that measured by {\it Planck} for local bright galaxies.  In the context of galaxy evolution models, this study confirms that galaxies with radio AGN also typically support hot gaseous halos. Adding {\it Herschel} observations allows us to show that the SZ signal is not significantly contaminated by dust emission.  Finally, we analyze the contribution of radio sources to the angular power spectrum of the cosmic microwave background.

\end{abstract} 


\section{Introduction}
\label{introduction}

In the standard cosmological structure formation scenario, astrophysical structures (galaxies, galaxy groups and clusters) form when gas condenses at the centers of hierarchically merging dark matter halos \citep{whiterees1978}.  The accretion is characterized according to two modes: radiatively efficient and radiatively inefficient \citep{silk1977, binney1977, reesandostriker1977}.  In the radiatively efficient rapid-cooling mode, the gas is shock-heated within the virial radius of the dark matter halo, and as it cools it accretes onto the galactic disk, contributing to star formation.  In the hydrostatic (hot) mode, the gas is shock heated before reaching the virial radius, and a hydrostatically supported gaseous halo forms.  This gaseous halo radiates in X-ray wavelengths, and as it radiatively cools, it has the capacity to form stars.  However, large cooling flows accompanied by prodigious star formation are not observed: the predominant stellar populations at the centers of the most massive halos are old and red, and X-ray measurements indicate a lack of cool gas compared to the level expected from the cooling \citep[e.g., ][]{oegerle2001, peterson2003}.   One mechanism invoked to explain this problem is feedback from active galactic nuclei (AGN).  Radiatively inefficient accretion can power a ``radio mode" AGN, where the gas that accretes onto the AGN ultimately comes from the hot gaseous halo.  Because the radio AGN can deposit energy back into this gaseous halo, a feedback mechanism can form that regulates the gas in a way that prevents both catastrophic cooling and excessive star formation.  Evidence for the existence of such feedback comes from a number of different observations \citep[for a review, see][]{mcnamara2007}.  Cavities or bubbles are observed in the X-ray halos of nearby galaxy clusters that are filled with radio emission \citep[e.g.,][]{fabian2000, mcnamara2000}. Galaxy evolution models that include this feedback mechanism can successfully reproduce observed galaxy luminosity functions \citep[e.g.,][]{bower2006, croton2006, sijacki2007}.  The correlation of the prevalence of low luminosity radio AGN with stellar mass, black hole mass, galaxy velocity dispersion and color also provides support for the presence of radio mode feedback \citep[e.g., ][]{best2005, best2012}.
That radio-loud AGN tend to reside in relatively massive ($\sim10^{13}$~M$_{\odot}h_{70}^{-1}$) dark matter halos has been confirmed by studies of the clustering of galaxies around radio AGN \citep{overzier03, magliocchetti04} and by stacked weak lensing \citep{mandelbaum09}.  However, the gaseous content is more challenging to observe, with recent X-ray studies currently restricted to  dozens of systems \citep{ineson2013}.

Sensitive millimeter-wave observations have the potential to reveal the Sunyaev-Zel'dovich effect of the hot 
gas associated with radio AGN.  The Sunyaev-Zel'dovich (SZ) effect is a spectral distortion in the cosmic microwave background (CMB) that occurs when CMB photons inverse-Compton scatter off of the hot ionized gas associated with dark matter halos.  The thermal SZ effect appears as a temperature decrement below (and increment above) $\sim$220~GHz, at which frequency there is no distortion \citep{sz70}.  The amplitude is effectively independent of redshift \citep[for a review, see ][]{carlstromreview}.  
Given the mass of the AGN-hosting dark matter halos, a simple extrapolation of SZ-mass relations derived from more massive halos \citep[e.g.,][]{sifon} reveals that the SZ effect in 1.4~GHz selected AGN should offset the non-thermal millimeter emission at 148~GHz by a significant ($\sim$50\%) fraction.  The \citet{plancklocalbrightgalaxies} recently reported a detection of the SZ effect of halos of similar mass that they selected via stellar mass. These observations suggest the feasibility of measuring the SZ effect from the halos that host radio-loud AGN with existing data sets. If the SZ effect can be separated from the synchrotron source spectral behavior, we can provide evidence of the existence of hot gas atmospheres in dark matter halos hosting galaxies with radio loud AGN.

At flux densities above 1~mJy, counts of 1.4~GHz radio sources are dominated by AGN that span a wide range in redshift, with median redshift $\sim1$ \citep{condon89}.  
The synchrotron spectral behavior of radio sources is complex and varies significantly from source to source \citep[e.g., ][]{dezotti10,sadler06,sajina10,lin2009}.  Models describing radio source spectral behavior often characterize sources into two types: flat-spectrum ($\alpha\sim0$, where $\alpha$ is defined such that $S \propto \nu^{-\alpha}$) and steep-spectrum ($\alpha\sim0.8$).  
Optically thin synchrotron emission, usually associated with radio lobes, results in a steep spectrum source, while optically thick emission, usually associated with compact cores, results in a flattened spectrum source \cite[e.g.,][]{dezotti10}. Using optical observations and radio data at multiple frequencies, the counts of 1.4~GHz sources have been successfully modeled with radio luminosity functions associated with these two populations  \citep[e.g.,][]{condon84,peacock85,danese87,dezotti05,massardi10}, with the flat spectrum population emerging as faint sources at $z=1$. These radio luminosity functions have been extrapolated, assuming a spectral index, to the millimeter regime where they have been used to predict the source population, especially for use in foreground models of CMB data \cite[CMB; e.g.,][]{toffolatti98,dezotti05,sehgal2010,tucci}. In addition to the synchrotron emission from the AGN, the AGN host galaxies can contain dusty star forming regions, as have been previously investigated by e.g., \citet{seymour2011}. With the advent of large scale millimeter-wave surveys with~mJy sensitivities and the availability of IR surveys with {\it Herschel}\footnote{{\it Herschel} is an ESA space observatory with science instruments provided by European-led Principal Investigator consortia and with important participation from NASA.} , we can now directly test the spectral behavior of 1.4~GHz-selected radio loud AGN across two decades of frequency.

In addition to the AGN population that dominates at $S_{1.4}>1$~mJy, deep radio surveys at 1.4~GHz (20 cm) \citep[e.g.,][]{condon84, becker95, condon95, hopkins98, mauch03} have revealed an excess number of sources at faint radio flux densities, typically with an upturn in the number counts below $\sim1$~mJy.  The observed excess has been mostly attributed to the emergence of a population of spiral galaxies undergoing significant star formation \citep[e.g., ][]{hopkins98, condon92}, although with some contribution from low-power AGN \citep[e.g.,][]{seymour08}. These star-forming galaxies (SFGs) are the predominant population below 0.1~mJy and account for $\ga20$\% of the population below $S\sim1$~mJy. 
  The radio emission from SFGs comes from synchrotron radiation associated with supernova remnants.  In the far-infrared (FIR), and similarly into the submillimeter regime, SFG emission is predominantly thermal from the absorption and re-emission of starlight by dust grains and is described as a gray body with a spectrum rising with frequency \citep[S $\sim \nu^{3.5}$; ][]{draine03}.  

In this study, we stack 148~GHz, 218~GHz and 277~GHz observations from the
Atacama Cosmology Telescope (ACT) corresponding to the locations of 1.4~GHz selected
radio sources 
to determine the average millimeter
emission associated with radio sources as a function of 1.4~GHz flux
density. 
We also utilize radio data from surveys conducted by Parkes and the Green Bank Telescope and infrared data from surveys conducted by {\it Herschel} to constrain the median spectral energy distributions (SEDs) of radio sources over the frequency range of 1.4~GHz up to 1200~GHz.
We use Faint Images of the Radio Sky at
Twenty-Centimeters (FIRST) sources matched to NRAO VLA Sky Survey (NVSS) sources through
the cross-matched catalogs provided by  \citet{best2012} and 
\citet{kimball08}. 
This enables us to benefit from both the $1''$ astrometric precision of the FIRST survey, and the ability of
the larger ($45''$ FWHM) NVSS beam to recover more extended radio emission.
By modeling the observed flux densities, we estimate the average spectral behavior of
low-frequency sources from 1.4~GHz to 1200~GHz and find a robust
detection of the SZ effect in radio-loud AGN halos, particularly evident at 148~GHz. 

This paper is organized as follows.  Section~\ref{sec:data} describes the data from ACT, the 1.4~GHz radio source samples, the IR data, and additional 5~GHz radio survey data used.   
Section~\ref{sedanalysis} presents the median SEDs for AGN and for SFGs from a relatively low redshift radio source sample constructed by \citet{best2012} by matching radio sources with galaxies with spectroscopic measurements from SDSS, as well as the results of modeling different components of these SEDs.  In Section~\ref{full_analysis} we extend this analysis to a larger sample of radio sources  from \citet{kimball08} that extends to higher redshift, albeit a sample that lacks optical counterpart identification, and split the sample into bins of 1.4~GHz flux density.  
Finally, we discuss and summarize our results and their implications for the SZ effect for the typical radio AGN halo and for the CMB power spectrum in Sections~\ref{discussion} and \ref{conclusions}. Unless specified otherwise, we assume a flat $\Lambda$CDM cosmology with $\Omega_m=0.30$, $\Omega_\Lambda=0.70$. For the Hubble constant, we adopt the notation $H_0=70h_{70}$~km\,s$^{-1}$\,Mpc$^{-1}$, and $E(z)$ describes the evolution of the Hubble parameter and is defined as $E(z) = \sqrt{\Omega_M(1+z)^3 + \Omega_{\Lambda}}$.  

\section{Data}
\label{sec:data}

\subsection{ACT Data}\label{act_data}

ACT is a six-meter, off-axis Gregorian telescope in the Atacama Desert of Chile at an altitude of 5200 meters \citep{swetz11}. The location was chosen for the atmospheric transparency at millimeter wavelengths and the ability to observe both northern and southern celestial hemispheres.  ACT had four observing seasons between 2007 and 2010 and surveyed both a southern celestial hemisphere region ($\delta = -53.5^{\circ}$) and an equatorial region. ACT observed simultaneously in three frequency bands centered on 148~GHz (2.0~mm), 218~GHz (1.4~mm) and 277~GHz (1.1~mm) with angular resolutions of 1.4$'$, 1.0$'$, and 0.9$'$, respectively.  Each band had a dedicated array of 1024 transition edge sensors.  This paper uses the 2009-2010 148~GHz, 2009-2010 218~GHz and 2010 277~GHz equatorial observations.   

Prior to all other analyses described in this paper, the ACT data were matched-filtered \citep[c.f.,][]{tegmark98} with the ACT beam \citep[][for more details see below in Section \ref{calibration_manifesto}]{hasselfieldbeam} in order to increase the signal to noise of compact sources. The filtering procedure is the same as described in \citet{marsden2013}.
Each map for each season was calibrated separately (for uncertainties in overall calibration, see below in Section \ref{calibration_manifesto}) and filtered with the beam appropriate for that season and band. The resulting calibrated, filtered maps were combined into a multi-season map via a weighted average, with the weights set for each pixel by the number of observations.  
The ACT sensitivity varies throughout the maps according to the depth of coverage.  The typical rms noise level is 2.2, 3.3 and 6.5~mJy for 148, 218 and 277~GHz, respectively.  

The ACT equatorial region contains detectable levels of Galactic cirrus emission.  In order to check whether Galactic contamination affects our results, we applied the dust mask used in the analysis of the angular power spectrum of the ACT data \citep{das}.  This mask was generated using data from the Infrared Astronomical Satellite (IRAS) as processed by \emph{IRIS} \citep{iris}.  Our results are not significantly affected when we mask the regions with strongest dust emission.

\subsubsection{Calibration}
\label{calibration_manifesto}

The uncertainty on the absolute temperature calibration of the 148~GHz band to WMAP is 2$\%$ \citep[at $l=700$;][]{sievers2013, 2011ApJ...740...86H}.  The relative calibration between the 148~GHz and 218~GHz bands was done by cross-correlation \citep{das}, and the resulting calibration uncertainty for the 218 GHz is 2.6$\%$. Therefore the 218~GHz calibration error is correlated with the 148~GHz calibration error. Because the CMB survey maps for the 277~GHz array have not yet been calibrated to WMAP, the 277~GHz calibration is derived from Saturn and Uranus observations \citep{hasselfieldbeam} and has a larger uncertainty of 7\%,  which is uncorrelated with the absolute calibration uncertainties of the 148~GHz and 218~GHz data. Additionally, errors in the assumed instrument beam and the map-making can propagate to uncertainty in the recovered ACT flux densities. In propagating these additional uncertainties, we follow the procedure used by \cite{marsden2013} and arrive at additional uncertainties in 148, 218, and 277~GHz flux densities of 1.9\%, 2.6\%, and 13\%, respectively.  The conservative 277~GHz flux density uncertainty reflects that this work presents the first reduction of these data, and we expect that as the reduction process for these data further matures, future work will have less uncertainty on the 277~GHz flux density scale.  

As discussed in \citet{dunner}, the ACT mapmaking pipeline recovers the flux densities of bright simulated sources to 1\%.  During the mapmaking procedure, bright sources are identified in a first-pass map and removed from the time-stream data before making the final map.  However, faint sources do not have this two-step procedure applied to them, so in principle their flux density recovery could differ from that of the bright sources.  We have tested the flux density recovery for faint sources by adding simulated sources to the data and putting them through the mapmaking procedures, and we find that the flux densities of faint ($<10$~mJy) sources are also recovered to 1\%. 

Combining the effects of calibration, beam measurement, and map-making, the flux density uncertainties for the 148~GHz band and  218~GHz band are  3\%, 5\%, respectively, and the correlated component is 3\%. The total  error for the 277 GHz band is 15\%, which is to a good approximation uncorrelated with flux density errors in the lower two ACT frequency bands. The full covariance is used when modeling flux density data in Sections~\ref{sedanalysis} and \ref{sec:modeling}.

\subsection{Radio Source Samples}
\label{first_sample}

The FIRST survey mapped over 10,000 deg$^{2}$ at 1.4~GHz (20 cm) from
1993 to 2004 using the VLA  
with imaging resolution of 5\arcsec \citep{becker95, white97}.  The resulting
source catalog has astrometric accuracy of 1$''$. The overall limiting
flux density threshold ranges from 0.75~mJy to 1~mJy in
its equatorial region. In addition to FIRST, the National Radio
Astronomy Observatory (NRAO) VLA Sky Survey \citep[NVSS; ][]{condon89}
was conducted at 1.4~GHz and covers the full sky above $-40^{\circ}$
declination to a depth of $\sim2.5$~mJy. 

\begin{figure}
	\centering
	\includegraphics[width=84mm]{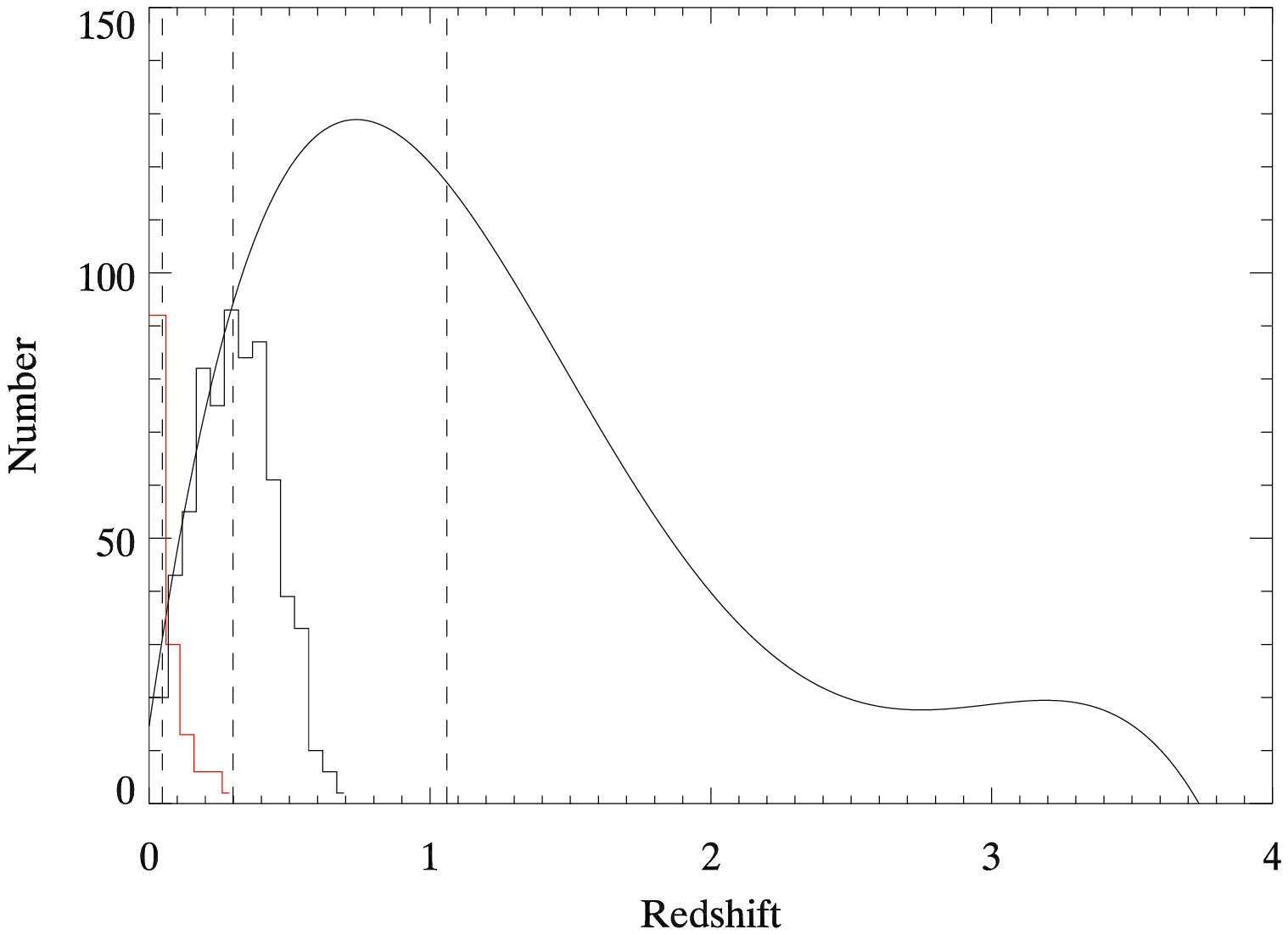}
\caption{The black histogram shows the redshift distribution of radio AGN from \citet{best2012}, with selection as described in Section \ref{first_sample}.  The red histogram shows the redshift distribution of SFGs from \citet{best2012}.  The black curve shows the redshift distribution we assume for the \citet{kimball08} sample, which is given in \citet{dezotti10} and based on data from \citet{brookes08}. The dashed lines indicate the median values for each sample.\label{fig:zdist}}
\end{figure}

 \citet{best2012} construct a sample of radio sources with spectroscopic redshifts by matching galaxies from the Sloan Digital Sky Survey (SDSS) with NVSS and FIRST.  The catalog has a limiting radio flux density of 5~mJy. The sample is split into star-forming galaxies and AGN according to radio and optical spectroscopic properties: the 4000{\AA}  break strengths and the ratio of radio luminosity to stellar mass, the ratio of radio to emission line luminosity, and a standard `BPT' emission-line diagnostic \citep{bpt, kauffmann2003}.  We use 667 radio-loud AGN in the region of overlap with the ACT survey (with median redshift of 0.30, see Figure \ref{fig:zdist}), and 149 star-forming galaxies (with median redshift of 0.05, see Figure \ref{fig:zdist}).  Only 27 of the 667 AGN are identified by \citet{best2012} as high excitation radio galaxies based on their optical emission lines, so the population can be modeled as dominated by low excitation radio galaxies.  We construct and analyze the SEDs of these populations in Section \ref{sedanalysis}.  

The \citet{best2012} sample has the twin advantages of providing redshifts and of differentiating between SFGs and AGNs.  It is, however, much smaller than samples drawn from radio surveys without optical IDs and confined to relatively low redshifts. To also exploit the full statistics of the sensitive radio surveys and to, albeit coarsely, investigate evolution with redshift, in Section \ref{full_analysis} we study a sample
drawn from a combined cross-matched
catalog\footnote[2]{\url{http://www.cv.nrao.edu/~akimball/radiocat/}} of
\citet{kimball08}. 
Because this sample is not restricted to radio sources matched to galaxies in SDSS, it extends to higher redshift. 
From this catalog, we use all FIRST detections falling within
the region of overlap with the ACT survey that are assigned at least one match to an NVSS
source. Matches are required to lie within
$30\arcsec$ of one another. 
Should there be multiple FIRST or NVSS sources within the $30\arcsec$ matching radius, we use
only the highest ranked match as per the matching algorithm detailed
in \citet{kimball08}. For the stacking analysis, we use the source locations from FIRST,
which are measured at high astrometric precision by the survey's $5\arcsec$
FWHM beam.
In our subsequent modeling of these sources, we use their associated NVSS
flux density measurements. The larger $45\arcsec$ FWHM beam of NVSS is
less prone to resolving out flux density from extended sources and thus
provides more accurate 1.4~GHz flux densities, as well as being a closer match to the ACT beam.
To avoid stacking on lobes of extended sources that are identified as separate sources in the FIRST catalog and more generally to avoid double counting sources that fall within a single ACT beam, we exclude sources that have neighboring sources within 1\arcmin. This leaves \TotalInRegionSources\  sources within the overlapping FIRST and ACT observing regions ($\sim$339 square degrees; $305^{\circ}\leq\alpha\leq58^{\circ}$, $-1.5^{\circ}\leq\delta\leq1.5^{\circ}$).  We restrict the sample to the \TotalInFluxRangeSources\ sources that have $5<S_{1.4}<200$~mJy.  
Because we model the average SED of these sources with an AGN model, we want to remove SFGs from the sample.  We mask any of these sources that lie within 5$'$ of SFGs from the catalog of \citet{best2012}.  This eliminates \BestSFGCutNum\ sources from the sample.  The expected number of SFGs for this sample is 
140 (for comparison, 149 SFGs were identified in \citealp{best2012}), as calculated using 1.4~GHz count models based on luminosity functions from \citet{dunlop90} and \cite{sadler02}.  For the remaining sources, assumed to be radio AGN, we adopt the redshift distribution from \citet{dezotti10} \citep[which was fit to data from][]{brookes08} when modeling the SED.  This distribution is shown in Figure \ref{fig:zdist} and has a median redshift of 1.06 (compared to 0.30 for the \citealp{best2012} sample).

Sources with ACT flux density greater than three times the rms noise level
of the map in any of the three bands are identified and excluded from the
stacking analyses.  Because the noise is determined locally, the flux density at which these cuts are drawn varies with position in the map, but typical values for the flux density threshold are 6~mJy at 148~GHz, 9~mJy at 218~GHz, and 18~mJy at 277~GHz. 
These sources represent outlier objects that are likely
bright based on orientation, such as blazars \citep{urry} or based on
chance alignments, such as rare, lensed star forming galaxies
\citep{negrello2010, vieira10, marsden2013}.  By removing these millimeter bright sources, we
reduce known inclination and lensing dependent selection effects
present in a small subset of the sources.  Similarly, negative $3\sigma$ deviations from the mean flux density in either band
are also excluded, which will exclude large galaxy clusters with SZ decrements at 148~GHz and reduce the bias potentially introduced by excluding positive deviations in flux density from the mean. 
These cuts based on the ACT flux densities exclude \ACTExcluded\ (\ACTExcludedPercent) 
radio sources from the stacking analysis of the \citet{kimball08} sample and 23 ($3.3\%$) AGN from the \citet{best2012} sample.

The synchrotron emission from radio-loud AGN is known to be variable.  The FIRST, NVSS and ACT surveys were not simultaneous, so variability could affect the inferred spectral indices.  Because the radio sources were selected based on their 1.4~GHz flux densities, variability would preferentially bias the 1.4~GHz flux density high relative to the 148 and 218~GHz flux densities, effectively steepening the average spectral index.   However, the 1.4~GHz selected sources do not tend to vary as much as sources selected at ACT frequencies \citep[which are more likely to be highly variable blazars, e.g., ][]{marriage11}.      
For example, \citet{thyagarajan2010} look for variability within the FIRST data,   using three potential criteria to determine variability: the distribution of the peak flux density at different times deviates significantly ($>5\sigma$) from a normal distribution, the maximum deviation of the peak flux density from the mean exceeds $5\sigma$, and the largest variation between data points on a light curve exceeds $6\sigma$.  They find 
 that for FIRST radio sources with counterparts identified as SDSS galaxies, the fraction that is variable at 1.4~GHz is $0.6\%$; the corresponding fraction for FIRST sources matched to SDSS quasars is 1$\%$.  
 
\subsection{Infrared data}
We investigate the ensemble SED properties of these AGN by calculating the median flux densities at their positions across a wide range of multi-wavelength data sets.  To investigate the contribution of dust to the SED of the radio sources, we take advantage of  surveys conducted by the {\it Herschel Space Observatory} \citep{herschel} using the Spectral and Photometric Imaging REceiver  instrument \citep[SPIRE;][]{spire} that overlap with the ACT survey region: HerMES Large-Mode Survey (HeLMS), which is part of the Herschel Multi-Tiered Extragalactic Survey \citep[HerMES; ][]{hermes}, and the publicly available Herschel Stripe 82 Survey\footnote[3]{http://www.astro.caltech.edu/hers/}  \citep[HerS;][]{hers}.

SPIRE has three bands, centered at approximately 500, 350 and 250~$\micro$m (corresponding to 600, 857 and 1200~GHz, respectively).  The maps used are made with SANEPIC \citep{patanchon2008}.
Of the radio loud AGN in the ACT region from \citet{best2012}, 384 (58\%) fall within the HerS or HerMES survey regions, as do 80 SFGs (54\%).  Of the radio sources in the sample from \citet{kimball08}, \SourcesWithIR\ fall within the HerS or HerMES survey regions.

\subsection{Flux density measurements}\label{sec:fluxdensity}

At 1.4~GHz, we use the cataloged NVSS flux densities corresponding to the \citet{best2012} and \citet{kimball08} sources.  Data from the Parkes-MIT-NRAO survey\footnote[4]{http://www.parkes.atnf.csiro.au/observing/databases/pmn/pmn.html} \citep[PMN;][]{pmn, pmnmaps} and a Green Bank Telescope survey \citep{gbsurvey} at 4.85~GHz provide an additional constraint on the ensemble radio spectral index.  
For PMN and GBT as well as the {\it Herschel} surveys, we measure the flux densities in the map of each survey at the source positions, which correspond to the positions of the optical counterparts for \citet{best2012} and to the positions of FIRST sources for \citet{kimball08}.  

For the ACT data, as discussed in \citet{marriage11}, given the form of the filter, the source-centered value of the filtered map multiplied by the solid angle of the beam is the source flux density. To calculate the flux densities from the map, ACT pixel values are corrected by a factor that accounts for averaging of the instrument
beam peak over the 0.5\arcmin-square pixel of the ACT maps.  This correction factor is applied to account
for the fact that measured flux density depends on the location of a source within a pixel. 
For example, a source will have a lower measured flux density if it is located at the junction of two pixels instead of
the center of a pixel. We make no effort to correct this mis-centering effect on a 
source-by-source basis, as the associated per-source dispersion is much lower than the rms
noise level and averages down to a negligible level in the  stack. Instead we apply an 
average correction given that any source has equal
probability of falling anywhere within the 0.5\arcmin-square pixel.
These factors correspond to \AROnePixwin~at 148~GHz, \ARTwoPixwin~at 218~GHz, and 
\ARThreePixwin~at 277~GHz.

We performed null tests by calculating the weighted average flux densities of randomly selected, source-free locations within 
the ACT$+$FIRST overlap region in all three ACT frequency bands.  These were computed in the 7 flux density bins used in Section \ref{full_analysis} for 1000 trials of \TotalStackedSources\ samples each. These null tests were found to be
consistent with no signal with a $\chi^2$ of
\NullChiSqAROne, \NullChiSqARTwo\ and \NullChiSqARThree\ at 148~GHz, 218~GHz and 277~GHz, respectively, each with \NullDOFAROne\ degrees
of freedom. The corresponding probability of a random realization to exceed the observed 148~GHz, 218~GHz and 277~GHz 
$\chi^2$ estimates are \NullPTEAROne, \NullPTEARTwo\ and \NullPTEARThree\ respectively.

Details on the stacking and uncertainty estimation for our analysis of the \citet{best2012} sample and of the \citet{kimball08} sample are provided in Sections \ref{sedanalysis} and \ref{full_analysis}, respectively.

\section{Spectral energy distribution construction and modeling}
\label{sedanalysis}

In this section, we investigate the median SED for the \cite{best2012} sample (with selection described in Section \ref{first_sample}), which has spectroscopic redshift measurements (median $z=0.30$) and was categorized into populations of AGN and SFGs. In this section, we model each of these populations independently.

\subsection{Stacked flux density measurements}

The SED models (described in Section \ref{sedmodel} and \ref{sfgmodel}) were fit to the median of the flux densities (calculated as described in Section \ref{sec:fluxdensity}) for each band.  The distribution of the flux densities of the sources (particularly at 1.4~GHz) is very broad, so we use the median flux densities in each band in order to lessen the influence of bright outliers.  For each source in the \citet{best2012} catalog, we use all data available. Some sources lie outside the footprints of the {\it Herschel}  surveys but fall within the ACT survey region, and we include their radio and millimeter flux densities in this analysis.  Thus the infrared median flux densities are drawn from a smaller sample of sources than the radio and millimeter median flux densities, but all sources are drawn from the same parent sample, with the IR sample restricted only by sky area, so selection differences should not be significant.  Indeed, the radio and millimeter median flux densities of the full sample are consistent within $1\sigma$ with the median flux densities of the subsample that lie within the {\it Herschel} survey areas.  The infrared surveys were treated as interchangeable, as the median flux densities for radio AGN in HerS are within the 1$\sigma$ uncertainties of the median flux densities of radio AGN in HerMES for each {\it Herschel} band.  

The covariance matrix for the median flux densities was constructed via a bootstrap resampling of the source flux densities.  Because the source flux densities are likely to be correlated between bands (a source bright in one Herschel band is likely to be bright in other Herschel bands), we retained off-diagonal elements of the covariance matrix.  
We have taken the overall photometric error for the SPIRE photometer to be $7\%$, of which $5\%$ is correlated between SPIRE bands.
We added the square of this photometric uncertainty to the variance of the {\it Herschel} bands and the square of the correlated component to the covariance between {\it Herschel} bands, and we similarly included the uncertainties in the ACT calibration described in Section \ref{calibration_manifesto} in the appropriate diagonal and off-diagonal terms of the covariance matrix. This covariance matrix was used to calculate the likelihood of the data given the model according to the following formulation:
\begin{equation}
-2 \ln \mathcal{L} \propto {\bf \it A}^{T} {\bf C}^{-1} {\bf \it A}
\end{equation}
where {\bf \it A} is equal to the data (median flux densities for each band) minus the model evaluated for a given set of parameters and {\bf C} is the covariance matrix (which does not depend on the model parameters). We used an MCMC analysis to determine the best fit parameters and associated uncertainties for the model.  

\subsection{Radio loud AGN}
\subsubsection{AGN spectral energy distribution model}\label{sedmodel}

We construct a model for the SED of the radio AGN containing components for the synchrotron, SZ effect and dust contributions.  The amount of free-free emission, which can contribute to the millimeter regime for star forming galaxies, is expected to be negligible, especially given how relatively little flux density it contributes to the spectral energy distributions of the much lower redshift star-forming galaxies in Section \ref{sfganalysis}.  
The synchrotron component is modeled with a single spectral index: $S\propto\nu^{-\alpha}$.  Although the spectral index may steepen due to electron aging, no deviation in $\alpha$ is significantly detected in any 1.4~GHz flux density bin our analysis of the larger sample (Section \ref{flatspectrumsources}).  The frequency dependence of the SZ spectral distortion of the CMB (for units of specific intensity) is given by the following equation (as in \citealp{carlstromreview}, with derivation from \citealp{sz70,sz72}):
\begin{equation}
g(x) = \frac{x^4 e^x}{(e^x - 1)^2} \left(x \frac{e^x + 1}{e^x - 1} - 4\right) (1 + \delta_{SZ}(x,T_e)) \label{szspec}
\end{equation}
where the dimensionless frequency $x = \frac{h \nu}{k_B T_{CMB}}$ and $\delta_{SZ}(x,T_e)$ is the relativistic correction to the frequency dependence.  The relativistic correction is typically only appreciable for massive galaxy
clusters, which comprise only a small fraction of the host halos of the AGN.  For comparison, the ACT SZ-selected cluster sample analysis \citep{hasselfieldclusters} uses relativistic corrections that range from $3-10\%$ of the cluster $y$ parameter for a sample of much more massive clusters (by roughly 2 orders of magnitude, as seen in Figure \ref{fig:comparemasses}) than is typical for this study.

The SZ spectral distortion described in Equation \ref{szspec} is attributed to the hydrostatically supported ionized gaseous halo around the AGN.   There may be an additional SZ signal associated with a non-radiating relativistic plasma inside of ``cocoons'' (e.g., observed as X-ray cavities in galaxy clusters) formed by the AGN during radio-mode feedback. The hypothesis of such a relativistic plasma is motivated by the fact that the minimum non-thermal pressure associated with the relativistic electrons sourcing the observed synchrotron from the cocoons is estimated to be an order of magnitude too small to be in hydrostatic equilibrium with the surrounding X-ray emitting gas \citep[e.g.,][]{blanton2002,ito2008}. The spatial extent of the cocoon ($R<50$~kpc) is significantly smaller than the extent of the ionized gaseous halo, but the cocoon's gas pressure will likely be on average higher and the spectral distortion different from that of the non-relativistic SZ effect of the larger host halo. The formation of a cocoon may be modeled analytically as a point explosion \citep{ostriker1988} using self-similar solutions to describe shock motion and associated gas state parameters \citep{sedov}. With this or other models of the cocoons, the spectral distortions of the CMB by the associated relativistic plasma \citep{wright1979} inside the cocoons can then be calculated \citep[e.g.,][]{yamada1999,platania2002,pfrommer2005,chatterjee2007}. These calculations, together with simulations \citep[e.g.,][]{chatterjee2008,scannapieco2008,prokhorov2010,prokhorov2012}, imply a signal in many systems that falls significantly below that of the SZ effect from the bulk of the non-relativistic gas of the halo and below the sensitivity of the current data. \citet{prokhorov2010} suggest that high pressure cocoons in high-redshift systems could produce an SZ effect at 217~GHz corresponding to flux densities greater than 1~mJy. Our results rule out the possibility that such high pressure systems characterize the average SED of radio loud AGN. Beyond this observation, we do not attempt to put constraints on SZ from the hypothesized non-radiating relativistic plasma inside cocoons and instead attribute all the SZ effect in our model to the non-relativistic gaseous atmospheres. As shown in Section \ref{agnsedresults}, this model is a good fit to the data given their current precision.

We parameterize our model using the measured amplitude of the SZ effect at 148~GHz, which we call $A_{SZ}$.  In order to use the same parameter to model the 277~GHz data, we take into account the SZ effect frequency dependence defined according to Equation \ref{szspec} as well as apply a beam (and therefore band) dependent correction factor that arises due to the effect of the filter transfer function.    
This correction factor is discussed in detail in 
Appendix \ref{scalingforprofile}.  For the \citet{best2012} sample, we calculate this correction factor to be \AROneToARThreeMedCorrBest\  (for comparison, the effective uncertainty on the calibration for the 277~GHz data is 15\%).   

The dust contribution is modeled by calculating the median flux density for the dust components of the source sample, with every source being assigned a model dust spectrum of the form $\nu^{\beta}B_{\nu}(T)$.  Each gray body spectrum corresponds to a single AGN, and the redshift is assigned according to the AGN's optical spectroscopic redshift. Thus the flux density observed at frequency $\nu$ for a source at redshift $z$ is given by 
\begin{equation}
S_{\nu}(L, z) = \frac{L_{IR}}{4 \pi (1+z) d_{M}^2(z)} \frac{(\nu (1+z))^{\beta} B_{\nu (1+z)}(T)}{\int (\nu')^{\beta} B_{\nu '}(T) d\nu '} \label{graybody}
\end{equation}
where the limits of the integral are 300 to 2.1$\times 10^4$~GHz, $d_{M}(z)$ is the comoving distance and $B_{\nu}(T)$ is the black body spectrum for a given dust temperature.  This parameterization is similar to that used in \citet{addison2012}, except that our model only includes a single temperature dust component. 
Each source is assigned a temperature of 20~K and an emissivity index of $\beta=1.8$, and we also investigate the effects of varying T and $\beta$ in Sections \ref{agnsedresults} and \ref{sec:modeling}.  \citet{hardcastle2013} fit dust SEDs to the average {\it Herschel} flux densities for a similarly selected sample of radio-loud AGN with SDSS spectra available.  They found the best-fit $\beta=1.8$.  With this $\beta$, for a model with a single temperature dust component, they found the best fit $T = 20.3$~K, although they also found evidence for multiple dust temperature components contributing.  We fit for a single parameter describing the amplitude of every spectrum, which we call $L_{IR}$, corresponding to the model's infrared bolometric luminosity of the dust component.  If there is a positive correlation between the AGN and dusty IR sources, there may be multiple dusty sources preferentially clustered near AGN that could contribute to the median flux density measured, potentially biasing it higher than for a single source \citep[as discussed for galaxies that contribute to the CIB; e.g., ][]{dole2006,marsden2009,viero2013}.  Physically correlated structures must be near the AGN in redshift space, so the spectral shape is likely to be unaffected, although the bolometric luminosity of the dust measured would become an upper limit for any single component.

 \subsubsection{Results of fitting the median AGN spectral energy distribution}\label{agnsedresults}
 
\begin{figure}
	\centering
	\includegraphics[width=84mm]{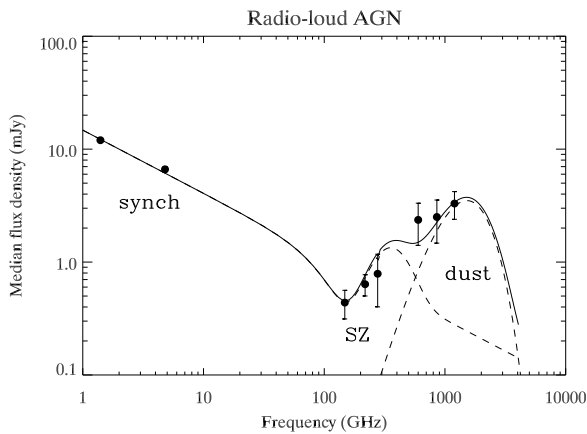}
\caption{The median flux densities in NVSS, PMN/GBT, ACT and {\it Herschel} surveys of radio AGN from \citet{best2012}.  As described in Section \ref{sedmodel}, a model with synchrotron, SZ effect and dust components is fit to the AGN data, and the model parameters describing the best fit (shown here) are listed in Table \ref{bestsedparameters}.  The dashed lines illustrate the synchrotron plus SZ effect component of the model and the dust gray body component of the model, and the solid line illustrates the full model evaluated for the best fit parameters.  The error bars on the medians shown are the diagonal elements of the covariance matrix computed via bootstrap sampling (off-diagonal elements were included for the fitting). Some of the error bars are smaller than the size of the symbols.  The $\chi^2$ of the fit is 5.8, with 4 degrees of freedom.  \label{fig:sed}}
\end{figure}

For the radio-loud AGN, we fit for the following parameters of the model outlined in Section \ref{sedmodel}: the radio spectral index ($\alpha$), the amplitude of the SZ effect at 148~GHz ($A_{SZ}$), the typical bolometric IR luminosity of the dust component expressed in solar units ($\log_{10}(L_{IR}/L_{\odot})$), and a normalization parameter for the synchrotron emission.  The best fit parameters are listed in Table \ref{bestsedparameters}.   The uncertainties quoted correspond to the $68\%$ confidence intervals returned by the MCMC for each parameter.  Figure \ref{fig:sed} shows the median flux densities for each band with the best-fit model.  The $\chi^2$ for the best fit AGN model is 5.8, with 4 degrees of freedom.  

\begin{figure}
	\centering
	\includegraphics[width=84mm]{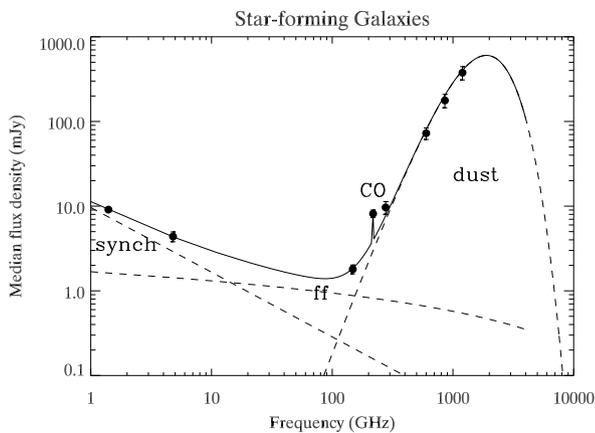}
         \caption{The median flux densities in NVSS, PMN/GBT, ACT and {\it Herschel} surveys of SFGs from \citet{best2012}.  Described in Section \ref{sfgmodel}, a model with synchrotron, free-free, the CO line for $z=0.05$ (rest frame frequency 230.5~GHz) and dust gray body emission is fit to the SFG data, and the model parameters describing the best fit (shown here) are listed in Table \ref{sfgparameters}.  The dashed lines illustrate the synchrotron component, the free-free emission component and the dust gray body component of the model, and the solid line illustrates the full model evaluated for the best fit parameters.  The error bars on the medians shown are the diagonal elements of the covariance matrix computed via bootstrap sampling (off-diagonal elements were included for the fitting). The error bars on the low frequency median flux densities are smaller than the size of the symbols.   
\label{fig:sfgsed}}
\end{figure}

\begin{table}
\caption{Best fit parameters for median AGN SED}
\begin{tabular}{@{}cc}
\hline
$\alpha $   &  $0.55 \pm 0.03$ \\
$A_{SZ} $ &  $0.45 \pm 0.13$~mJy \\
$\log_{10}(L_{IR}/L_{\odot}) $ &  8.7 +0.1/-0.3\\
$S_{sync}$ & $12.2 \pm 0.5$ mJy\\
\hline
\end{tabular}
\label{bestsedparameters}
\end{table}

\begin{table}
\caption{Best fit parameters for median SFG SED}
\begin{tabular}{cc}
\hline
$\alpha $   &  0.9 +0.5/-0.4 \\
$S_{sync}$ & 5.5 +0.7/-0.6 mJy\\
$EM_{\mathit{ff}} $ &  943  +245/-308 cm$^{-6}$pc\\
$S_{CO}$ & $2.8 \pm0.8$ mJy\\
$\log_{10}(L_{IR}/L_{\odot}) $ &  9.13 +0.07/-0.09\\
\hline
\end{tabular}
\label{sfgparameters}
\end{table}

As evident in Figure \ref{fig:sed}, the infrared data constrain the dust contribution to the millimeter region of the spectra to lie below the detected median ACT flux densities for the radio AGN.  There is $3\sigma$ level evidence that the SZ effect contributes to the shape of the spectrum in the ACT frequencies.  If we set the $A_{SZ}$ term to 0, the $\chi^2$ is 16.0, with 5 degrees of freedom.  
The value and significance of $A_{SZ}$ are robust to changes in the assumed dust temperature ($T=10, 15, 20, 30$~K were tested) and $\beta$ ($\beta=2.0, 1.8, 1.5, 1.0, 0.5$ were tested).  We reevaluated the best fit parameters for each set of T, $\beta$ listed and found that for any of the models within this parameterization, the best fit $A_{SZ}$ is within 0.5$\sigma$ of the quoted value.  The best fit value for $\log_{10}(L_{IR}/L_{\odot}) $ is not robust to changes in $\beta$ and T, but increases both with increasing $\beta$ and with increasing T, with T having a larger effect for the range of values probed.

We have also experimented with models that do not involve an SZ effect signal, but rather attempt to fit the data by altering source spectra. Unfortunately, few datasets constrain the spectral behavior of large samples of radio sources at these frequencies. The {\it Planck} mission constrains the millimeter-wavelength SEDs of very bright (for example, matched to sources with $S_{20}>300$~mJy) radio blazars \citep{planck_ercscvalid} and finds some evidence for spectral steepening.  Although we exclude such sources from this analysis, we nonetheless consider a model with a spectral index change of -0.5 at 70 GHz and allow the spectral index of the dust re-emission ($\beta$) to take on values that produces a minimum chi-square. For this model, $\beta = 0.9$ produces the best fit, but the resulting $\chi^2$ is 9, compared to the $\chi^2$ of 5.8 for the baseline model (with no spectral steepening, $\beta$ set to 1.8, and a term for the SZ effect amplitude). 
When $\beta$ is set to 1.8, which is typical for nearby galaxies \citep{smith2013}, the resulting $\chi^2$ is 11.  However, the synchrotron spectral behavior for the bright {\it Planck} sources, which consist mostly of blazars, differs significantly from the average spectral behavior of the sources in our sample.  To allow more flexibility in the synchrotron shape, we adopt a model where we fix the dust spectrum $\beta$ to be 1.8 and the SZ effect contribution to be 0, but introduce parameters for the location of the break in the synchrotron spectrum ($\nu_{break}$) and the amount of steepening ($\delta \alpha$, defined such that $\alpha(\nu>\nu_{break}) = \alpha + \delta \alpha$).
The best fit model values for $\alpha$, $\delta \alpha$, and $\nu_{break}$ are $0.47\pm 0.05$, 0.2 +0.3/-0.5, and $\sim$4~GHz, respectively. The location of the break, $\nu_{break}$, is not well constrained.  The $\chi^2$ value is 5.8, with 3 degrees of freedom.  If we reintroduce the SZ effect term, $A_{SZ}$, into the model and fix $\nu_{break}$ to be 5~GHz, the resulting best fit parameters for $\alpha$, $\delta \alpha$, and $A_{SZ}$ are $0.48\pm0.04$, $0.16\pm0.08$, and $0.3\pm0.15$, respectively.  The $\chi^2$ is 1.3, with 3 degrees of freedom.  Thus even with this flexibility, a model with SZ is preferred, although with much lower significance than for a model with a single synchrotron spectral index.  Although this model does provide a better fit than our fiducial model, the location of the spectral break at 5~GHz is not well motivated physically.  When we investigate this for the larger \citet{kimball08} sample in Section \ref{flatspectrumsources}, we again find that the model including the SZ effect is preferred, and more robustly.  For that sample, spectral steepening is not preferred, and a spectral break at 5~GHz is particularly unlikely (see Figure \ref{fig:model_distributions}).

 \subsection{SFG spectral energy distribution}\label{sfganalysis}
\subsubsection{SFG model}\label{sfgmodel}
The main components of the spectra of star forming galaxies include synchrotron emission that dominates at low frequencies, dust emission that dominates in the infrared, and free-free emission that contributes in the millimeter wavelength regime.
We adopt the same spectral model as used by \citet{peel2011}, who investigate the millimeter wavelength SED's of nearby late-type galaxies using data from {\it Planck}.  The synchrotron component is modeled as $S\propto \nu^{-\alpha}$, and the dust component is measured as a gray body spectrum with the same form as Equation \ref{graybody}. This model for the dust emission is supported by recent work: \citet{clemens2013} find that single temperature ($21$K) dust spectra (median fit $\beta=1.83$) adequately describe the millimeter wavelength through infrared SEDs of local star-forming galaxies, using $Planck$ High Frequency Instrument data in combination with IR data from {\it Wide-Field Infrared Survey Explorer} ({\it{}WISE}), {\it Spitzer}, {\it IRAS}, and {\it Herschel}. We fit for the parameters $\alpha$ and $L_{IR}$.    In addition to the dust emission, the free-free emission is expected to contribute to the ACT bands and is modeled as 
\begin{equation}
S_{\mathit{ff}} = 2 \times 10^{26} k_B T_{e} e^{(1 - {\tau_{\mathit{ff}}})} \Omega \nu^2 / c^2 
\end{equation}
where $S_{\mathit{ff}}$ is expressed in Jy, $k_B$ is the Boltzmann constant in J/K, $T_e$ is the electron temperature in K, $\Omega$ is the solid angle the galaxy subtends in steradians, $\tau_{\mathit{ff}}$ is the optical depth, $\nu$ is the frequency in Hz and $c$ is the speed of light in m/s.  We assume $T_e = 8000$~K, as is typical for the Milky Way, and calculate $\Omega$ for a 10~kpc disk.  The optical depth is calculated by
\begin{equation}
\tau_{\mathit{ff}} = 3.014 \times 10^{-2} T_{e}^{-1.5} \nu^{-2} EM_{\mathit{ff}} g_{\mathit{ff}}
\end{equation}
where $\nu$ is expressed in GHz, $T_{e}$ in K, $EM_{\mathit{ff}}$ is the emission measure in the commonly cited units of cm$^{-6}$pc and $g_{\mathit{ff}}$ is the Gaunt factor, which causes curvature in the high frequency free-free spectrum and is approximated as $g_{\mathit{ff}} = \ln(4.955 \times 10^{-2} / \nu) + 1.5 \ln(T_e)$ for $\nu$ in GHz and $T_{e}$ in K.  The numerical factors are given by \citet{oster1961}.  We fit for the emission measure to characterize the contribution of the free-free emission to the SFG spectrum.

In addition to these sources of continuum emission, we expect that the 218~GHz band will contain some contribution from the CO~$J(2-1)$ spectral line at 230.5~GHz.  The redshift distribution of the SFGs spans from 0.010 to 0.283 (see Figure \ref{fig:zdist}), with median redshift of 0.047.  The ACT 218~GHz band is 17.0~GHz wide, with central frequency at 219.7~GHz \citep{swetz11}.  Approximating the band transmission as a step function, for sources in the range $0.01<z<0.09$, the CO line will fall within the ACT band.  This corresponds to $80\%$ of the SFGs in the sample that fall within the ACT survey region.  The true contribution of the CO line to the average SED depends on the detailed shape of the band transmission.  In order to include the CO line in the SED model, we have added a parameter for additional flux density in the 218~GHz band.  As a result, the 218~GHz data do not constrain the SED continuum, although obtaining a reasonable value for the CO flux density contributed can indicate consistency.  

\subsubsection{SFG modeling results}
The SED model was fit to the median of the flux densities of the SFGs for each band as described for the AGN in Section \ref{sedmodel}. We fit for the following parameters of the model outlined above: the radio synchrotron spectral index ($\alpha$), a normalization parameter for the synchrotron emission at 1.4~GHz, the emission measure of the free-free emission ($EM_{\mathit{ff}}$), the additional flux density at 218~GHz attributed to CO line emission ($S_{CO}$), and the typical bolometric IR luminosity of the dust component expressed in solar units ($\log_{10}(L_{IR}/L_{\odot})$). Best-fit parameters are listed in Table \ref{sfgparameters}, and the best-fit model is shown with the median flux densities in Figure \ref{fig:sfgsed}.  The $\chi^2$ for the best fit SFG model is 4.5, with 3 degrees of freedom. As with the AGN model, the covariance includes both measurement uncertainty and intrinsic variation in the flux densities.

We can compare the best fit values of the median SFG SED to published models for nearby star-forming galaxies in the literature.  \citet{peel2011} use $Planck$ data to constrain the SEDs of M82, NGC~253 and NGC~4945.  They fit for more parameters than our data can constrain (particularly for the gray body spectrum, such as the dust temperature and $\beta$).  Comparing with their results provides a consistency check. 
 The emission measures they quote for their best fit model are $920\pm110$~cm$^{-6}$pc for M82, $284\pm17$~cm$^{-6}$pc for NGC~253 and $492\pm81$~cm$^{-6}$pc for NGC~4945.  Our ensemble value of 943  +245/-308~cm$^{-6}$pc is in agreement with M82.  Their best fit synchrotron spectral indices range from $1.1\pm0.1$ to $1.6\pm0.4$, and our value of $0.9+0.5/-0.4$ also agrees well.  Degeneracy between $\alpha$ and $EM_{\mathit{ff}}$ is expected such that steeper values of $\alpha$ imply higher values of $EM_{\mathit{ff}}$, and this degeneracy is observed in the MCMC posterior distribution of those parameters.  Including data for more bands at low frequency would better constrain these parameters.  

We can also compare the CO line flux density with expectations from the literature.  \citet{bayet2006} measure the flux density of the CO~$J(2-1$) line for nearby starburst NGC~253.  
Scaling by the square of the luminosity distances, the CO~$J(2-1)$ flux density at the median SFG redshift would be 36~mJy.  NGC~253 is a bright starburst, so its CO line flux density is likely brighter than that of typical SFGs.  The gas density of star forming galaxies correlates with the star formation rate through the well-known Kennicutt-Schmidt relation \citep{schmidt1959,kennicutt1998}. \citet{genzel2010} find a linear relation between the CO luminosity and the far IR luminosity for star forming galaxies.  If we scale the flux density by the ratio of the bolometric luminosity of NGC~253 \citep{rice1988} to the bolometric IR luminosity best fit to the SFG median SED, the expected CO flux density implied is 3.6~mJy, which lies within 1$\sigma$ of the best fit CO flux density contribution to the SED.  

\section{Millimeter wavelength behavior for different radio source flux densities}\label{full_analysis}
 
In this section, we investigate how the SEDs of the radio sources
relate to their 1.4~GHz radio source flux density by binning the sample from \citet{kimball08}.  This sample is much larger than that used in Section \ref{sedanalysis}, but it lacks optical counterparts and thus lacks individually measured spectroscopic redshifts and classification as AGN or SFGs.  This sample also extends to higher redshift (Figure \ref{fig:zdist}), enabling a comparison of properties of a low redshift sample (discussed in Section \ref{sedanalysis}) and a higher redshift sample.

 \subsection{Stacked flux density measurements}
 
\begin{table*}
\begin{minipage}{165mm}
\caption{Median flux density at radio source locations.}
\begin{tabular}{@{}cccccccc|ccc}
\hline
    Bin &
    $S_{1.4}$ Range &
    $N_{bin}$ &
    $S_{1.4}$ &
    $S_{4.8}$ &
    $S_{148}$ &
    $S_{218}$ &
    $S_{277}$ &
    $S_{600}^\dagger$ &
    $S_{857}^\dagger$ &
    $S_{1200}^\dagger$ \\
         &
        (mJy)&
        &
        (mJy) &
        (mJy) &
        (mJy) &
        (mJy) &
        (mJy) &
        (mJy) &
        (mJy) &
        (mJy) \\
\hline
 1 & 5.00 -- 8.47 & 1767 &  6.43 &  3.7 $\pm$  0.2 & 0.37 $\pm$ 0.05 & 0.66 $\pm$ 0.08 &  1.0 $\pm$  0.2 & \multirow{7}{*}{ 3.9 $\pm$  0.4} & \multirow{7}{*}{ 4.4 $\pm$  0.4} & \multirow{7}{*}{ 4.4 $\pm$  0.4} \\
 2 & 8.47 -- 14.3 & 1092 &  10.9 &  5.7 $\pm$  0.3 & 0.38 $\pm$ 0.07 & 0.79 $\pm$  0.1 &  1.3 $\pm$  0.2 &  &  &   \\
 3 & 14.3 -- 24.3 & 672 &  18.5 &  7.8 $\pm$  0.3 & 0.59 $\pm$ 0.09 & 0.83 $\pm$  0.1 &  1.5 $\pm$  0.3 &  &  &   \\
 4 & 24.3 -- 41.2 & 412 &  31.3 &  12. $\pm$  0.5 & 0.64 $\pm$  0.1 &  1.1 $\pm$  0.2 & 0.96 $\pm$  0.3 &  &  &   \\
 5 & 41.2 -- 69.7 & 222 &  53.5 &  21. $\pm$  0.9 & 0.94 $\pm$  0.1 &  1.4 $\pm$  0.2 &  1.5 $\pm$  0.5 &  &  &   \\
 6 & 69.7 -- 118. & 118 &  88.4 &  30. $\pm$   1. &  1.4 $\pm$  0.2 &  1.8 $\pm$  0.3 &  2.7 $\pm$  0.7 &  &  &   \\
 7 & 118. -- 200. & 61 &  149. &  45. $\pm$   1. &  1.9 $\pm$  0.3 & 0.83 $\pm$  0.4 &  1.5 $\pm$  0.9 &  &  &   \\
 \\
\hline
\end{tabular}
\medskip
{$\dagger$}{The {\it Herschel} data shown correspond to a single 1.4~GHz flux density bin containing all radio sources in the sample within the {\it Herschel} survey area. }
\label{first_results}
\end{minipage}
\end{table*}

\begin{figure*}
	\centering
	\includegraphics[scale=.95]{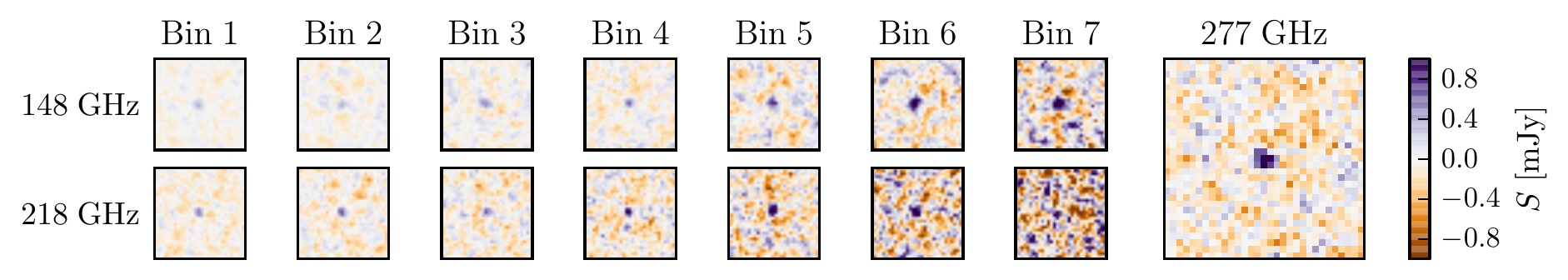}
	\caption{Thumbnail images (0.25$^{\circ}\times$0.25$^{\circ}$) of filtered ACT data stacked on FIRST source locations.  While the 148~GHz sources fade in the lower flux density bin, the 218~GHz sources remain visible.  The 277~GHz map is noisier, so we have shown the 277~GHz flux densities stacked for the entire sample instead of splitting it into 1.4~GHz flux bins, although we evaluate the 277~GHz stacked flux densities for each bin for use in the analysis in Section \ref{sec:modeling}.
	The 1.4~GHz flux density bins used are listed in Table \ref{first_results}.   \label{fig:first_submaps}}	
\end{figure*}

We compute the mean of the flux densities, which are calculated as described in Section \ref{sec:fluxdensity}, corresponding to subsets of
sources from the \citet{kimball08} radio source catalog (with selection described in Section \ref{first_sample}).  First, we logarithmically bin sources by their associated
1.4~GHz NVSS flux density ($S_{1.4}$) into \NBins\ bins with centers ranging from \LowBinFlux\ to \HiBinFlux\ ~mJy.  
Then, for each source within a flux density bin,
we determine the ACT flux density and the number of ACT observations
at the FIRST position of the source, which determines the noise in any given pixel \citep[see][]{marriage11}. Weighted by the number of observations, 
the ACT 148~GHz, 218~GHz and 277~GHz flux densities ($S_{148}$, $S_{218}$, and $S_{277}$ respectively) 
are then averaged to give the stacked ACT flux density for each 1.4~GHz flux density bin.  Weighted averages are similarly calculated for the GBT/PMN data and {\it Herschel}, with the weights defined as the inverse of the square of the uncertainties on the flux density measurements.  For the {\it Herschel} surveys, which do not cover the entire region of sky used in the ACT analysis (\SourcesWithIR\ out of \TotalStackedSources\ of the radio sources used in this analysis lie within Herschel survey footprints), we combine all of the sources across 1.4~GHz flux density bins, calculating a single ensemble averaged flux density for each {\it Herschel} band.  Constraining a single dust component also has the benefit that we do not need to know the redshift distribution of sources as a function of their 1.4~GHz flux densities and can model the full population as has been studied in the literature.

The distribution of ACT stacked flux densities within each bin was investigated
via a Monte Carlo bootstrap analysis, resampling the flux density distribution with replacement for 1000 trials. 
We use the weighted averages of the observed sample flux densities, and we adopt the uncertainties based on the bootstrap samples.  For the modeling in Section \ref{full_analysis}, we only include the covariance from the calibration uncertainties calculated in Section \ref{calibration_manifesto} and not bootstrap sampled covariance because we bin by 1.4~GHz flux density, which limits the amplitude of the cross-frequency-band sample covariance relative to the dominant noise in the map (unlike in Section \ref{sedmodel}, which includes sources across the full range of 1.4~GHz flux density).  

\begin{figure}
	\centering
	\includegraphics[width=84mm]{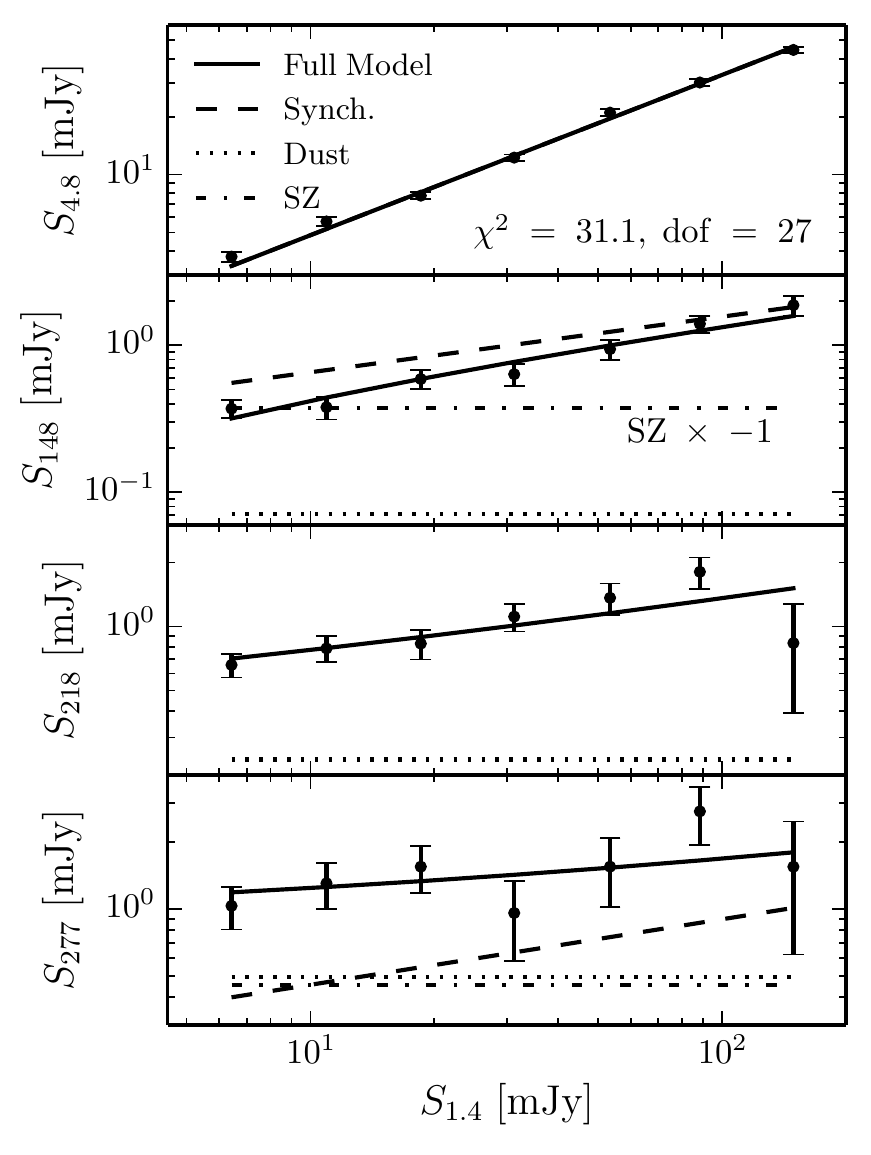}
\caption{The PMN/GBT stacked flux densities at 4.8~GHz and the ACT stacked flux densities at 148~GHz, 218~GHz and 277~GHz, with the curves corresponding to the best-fit model and its components overlaid. The dashed line indicates the AGN emission and the dotted line indicates the contribution from dust emission. The dot-dashed line indicates SZ effect of the halos hosting the AGN ($A_{SZ}$), which is negative for $\nu<218$~GHz. The solid line indicates the sum of the AGN emission, the SZ effect and the dust contribution.  All data are listed in Table~\ref{first_results}.\label{fig:first_stacked_all}}
\end{figure}

Table~\ref{first_results} lists the results of stacking the PMN/GBT, ACT and Herschel data at the FIRST source locations for the radio source sample from \citet{kimball08}.  
Thumbnails of the stacked ACT data for each 1.4~GHz flux density bin of the \citet{kimball08} sample are shown in Figure~\ref{fig:first_submaps}.  
Figure~\ref{fig:first_stacked_all} shows the stacked ACT flux densities at 148~GHz, 218~GHz and 277~GHz. 
The stacked ACT flux densities associated with the FIRST sources are detected at $\geq 3\sigma$ significance at 148 and 218~GHz for all but the highest 1.4~GHz flux density bin.  
The stacked ACT flux densities typically lie between 0.2 and 2.0~mJy.  The faintest 1.4~GHz flux density bin contains \LowBinCount\ sources and the brightest contains just \HiBinCount\ sources.

The average $S_{148}$, $S_{218}$ and $S_{277}$ are always lower than the corresponding $S_{1.4}$, as would be expected for synchrotron dominated sources with radio spectral index $\alpha_{1.4-148}>0$, where flux density $S \propto \nu^{-\alpha}$.   However, the average $S_{218}$ and $S_{277}$ are greater than the average $S_{148}$ for all but the highest corresponding bin in $S_{1.4}$, and for most bins $S_{277}$ is also higher than $S_{218}$.  In contrast, the synchrotron sources detected in ACT and South Pole Telescope (SPT) data, which are typically blazars and would be excluded from this analysis, tend to have flat spectral indices ($\sim0.2$) from low frequency (1.4~GHz or 5~GHz) to the millimeter regime, but then have falling spectral indices ($\alpha \sim 0.5$) between the millimeter bands \citep{vieira10, marsden2013, mocanu2013}.  There are two possible explanations for this spectral behavior in the stacked ACT flux densities of radio sources: either the 1.4~GHz selected sources have, on average, an inverted millimeter spectral index $\alpha_{148-218} <0$ and $\alpha_{148-277}<0$ or the SZ effect is causing the SED to rise from a decrement at 148~GHz through a null at 218~GHz to an increment at 277~GHz.  We investigate each of these scenarios in the following sections, finding that the SZ effect provides a better fit to the data and thus is the preferred explanation.

\subsection{Spectral Energy Distribution Modeling}
\label{sec:modeling}

We fit the model outlined in Section \ref{sedmodel} to the weighted average flux densities for each 1.4~GHz flux density bin.  
For this model we allow the synchrotron spectral index $\alpha$ to vary as a function of 1.4~GHz flux density with the following parameterization:
\begin{equation}\label{alphaequation}
       \alpha = \alpha_0 + \alpha_1 \log_{10}(\langle{S_{1.4}}\rangle/S_{pivot})) 
\end{equation}
and where $\alpha_0$ and $\alpha_1$ are fit as a function of $S_{1.4}$.  The central frequencies used here correspond to the synchrotron effective band centers \citep{swetz11}. The $S_{1.4}$ normalization ($S_{pivot}=\SPivmJy$~mJy, corresponding to the average 1.4~GHz flux density for the highest bin) has been chosen to minimize degeneracy between $\alpha_0$ and $\alpha_1$.  Recalling that the AGN population can be characterized by a bimodal distribution including flat ($\alpha\sim0$) and steep ($\alpha\sim0.8$) spectrum sources and that the average synchrotron spectrum could also be affected by steepening due to electron aging, we also investigate whether there is evidence for a break in the average synchrotron spectrum (see Section \ref{flatspectrumsources}).  We parameterize the $A_{SZ}$ as constant across all 1.4~GHz flux density bins, motivated by previous observations that the distribution of radio luminosities does not correlate with host galaxy mass or cluster velocity dispersion \citep{best2005, best2007}.  

We adopt the redshift distribution based on data from \citet{brookes08} with the form given by \citet{dezotti10}. The redshift distribution enters the model in the construction of the composite gray body spectrum (Equation \ref{graybody}) and in the angular shape of the average SZ pressure profile, which affects how the SZ signal is sampled by the filter as first discussed in Section \ref{sedmodel} and further discussed in Appendix \ref{scalingforprofile}.  For this assumed redshift distribution, the measured SZ effect at 277~GHz is \AROneToARThreeMeanCorr\ lower than that inferred by the spectral behavior (Equation \ref{szspec}) and the SZ effect amplitude at 148~GHz.  This correction factor is included in the construction of the model, which is parameterized by the SZ effect amplitude at 148~GHz, $A_{SZ}$.  Similarly, because the redshift distribution for this sample differs from that of \citet{best2012}, the same intrinsic amplitude of the pressure profile would produce different measurements for $A_{SZ}$ due to the effect of the filtering on the pressure profiles.  We take this effect into account before comparing the integrated $Y$ parameter in Section \ref{szdiscussion}.

The CO~$J(2-1)$ spectral line at 230.5~GHz is not likely to contribute to the SED because for the vast majority of the sources, the redshift would place the CO line outside of the ACT band.

\begin{figure}
	\centering
	\includegraphics[width=84mm]{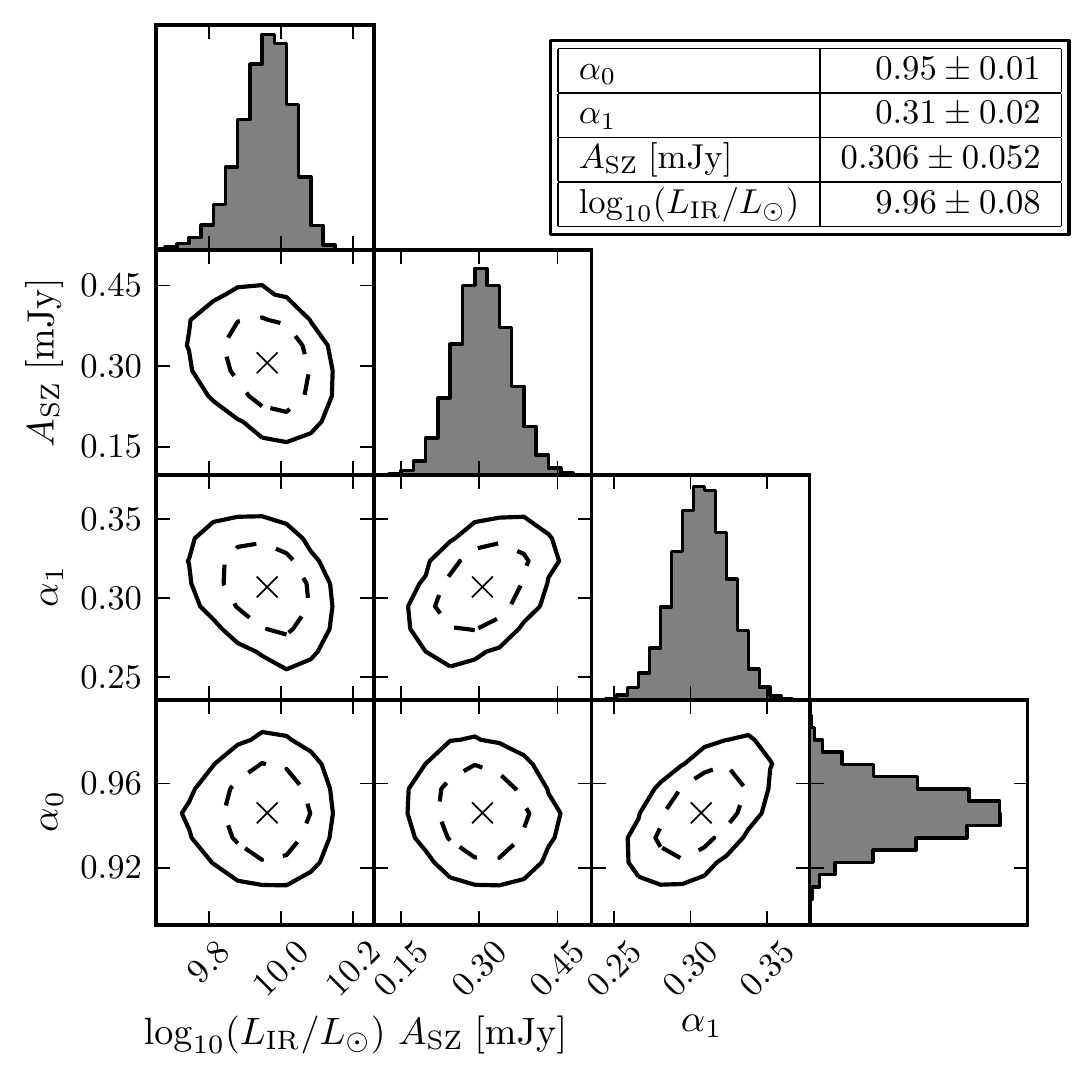}
\caption{The posterior distributions of the key model parameters, with best fit values listed. The histograms show the single-parameter distributions for each parameter. The dashed lines show the 68\% confidence regions, and the solid lines show the 95\% confidence regions. The parameters correspond to the effective AGN spectral index at the highest 1.4~GHz flux density ($\alpha_0$), how the AGN spectral index varies with 1.4~GHz flux density ($\alpha_1$), the amplitude of the SZ from ionized gas in AGN dark matter halos ($A_{SZ}$), and the bolometric IR luminosity ($\log_{10}({L_{IR}/L_{\odot}}$)). \label{fig:model_distributions}}
\end{figure}

We perform an MCMC to determine the posterior probability
 distributions for the parameters $\alpha_0, \alpha_1, A_{SZ}$ and $\log_{10}(L_{IR}/L_{\odot})$, which are given uniform priors.  We account for the full covariance of the uncertainties on the calibration and recovery of the ACT and {\it Herschel} flux densities (Section~\ref{calibration_manifesto}).  
Figure \ref{fig:first_stacked_all} shows the model evaluated for the best fitting parameters along with the data. Figure \ref{fig:model_distributions} shows the resulting parameter distributions for the main parameters of interest. All chains show good convergence, with Gelman-Rubin R - 1 parameter $< 0.005$.
  The best fit model has a $\chi^2=\ModelChiSqWithSZ$ with \ModelDOFWithSZ\ degrees of freedom (PTE=\ModelPTEWithSZ), so the model is a good fit to the data. 

Similarly to Section \ref{agnsedresults}, we vary the dust temperature, $T$, and emissivity, $\beta$, of the assumed graybody dust model to determine whether our choice of $T = 20$~K and $\beta = 1.8$ affect our best fit values. For every combination of $\beta=1.5, 1.8$ or $2.0$ and $T=10, 15$ or $20$~K, the constraints on all parameters, with the exception of $\log_{10}(L_{IR}/L_{\odot})$, are robust to changes in these assumption (varying at the $<1\sigma$ level). Moreover, the infrared data strongly disfavor dust models with lower dust temperature, such that the $\chi^2$ increases by $> 13$ for $T=15, 10$~K models relative to the $T=20$~K model for the ranges of $\beta$ tested.

\begin{figure}
	\centering
	\includegraphics[width=84mm]{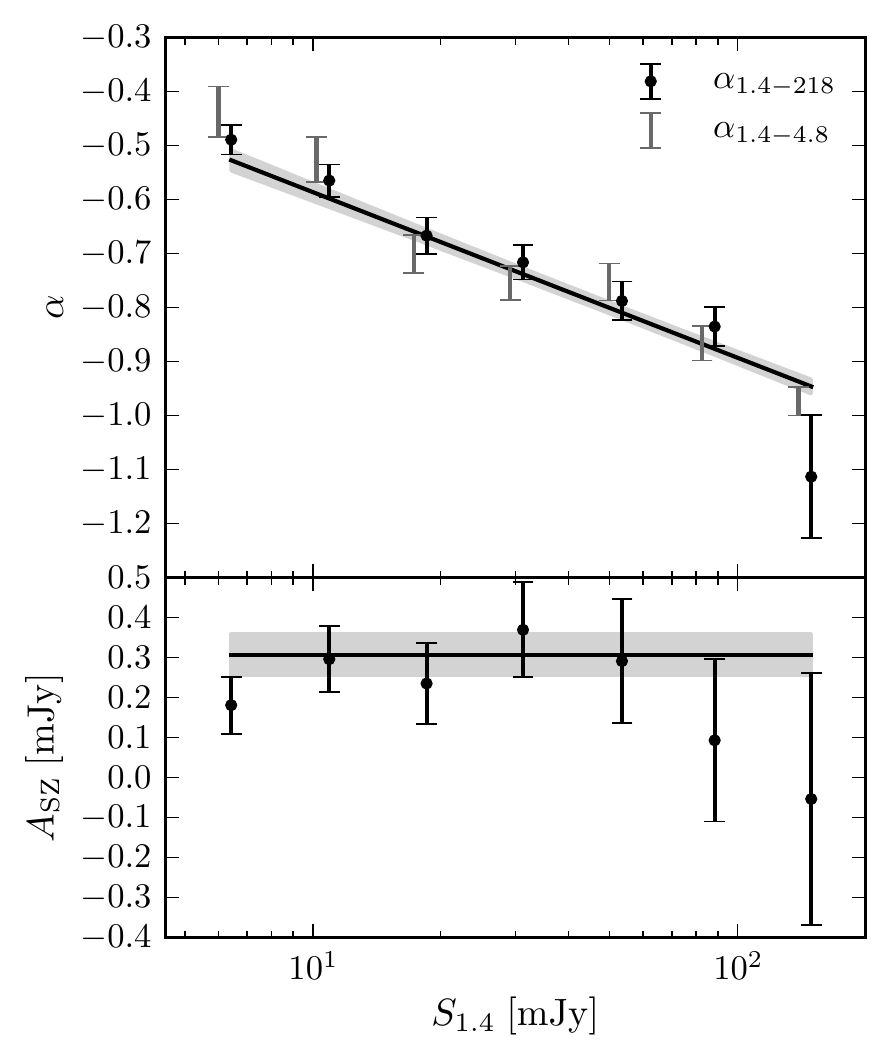}
\caption{Top: the effective 1.4~GHz - millimeter spectral index for each 1.4~GHz flux density bin (points), compared to the best-fit model (solid line), for which the slope and normalization are fit. Bottom: the amplitude of the term that models the SZ signal associated with AGN, compared to the best-fit model, which is constrained to be a constant with 1.4~GHz flux density. In both plots, the gray regions indicate the 1$\sigma$ uncertainty on the best-fit parameters of the model. For the top plot, each data point shown is calculated for a given 1.4~GHz flux density bin by solving $ \langle{S_{218}}\rangle  =  \langle{S_{1.4}}\rangle (217.6/1.4)^{-\alpha} $ for $\alpha$ (the gray points show the equivalent for 4.8~GHz instead of 217.6~GHz).  The data points shown in the bottom plot are calculated as the difference between the measured and expected $\langle{S_{148}}\rangle$.  For the model fit, we use the combined dataset with the appropriate covariances taken into account.  
 \label{fig:uberspec}}
\end{figure}

The parameter estimates from this analysis shed
light on AGN and the SZ effect from radio-loud AGN hosts.  Figure \ref{fig:uberspec} shows the best fit model for $\alpha$ (Equation \ref{alphaequation}) and the amplitude of the SZ effect, along with the measured values for each 1.4~GHz flux density bin. The SZ effect associated with AGN hosts
$A_{SZ}=$ \ParASZWithSZ~mJy is detected with \ParASZWithSZSignif~$\sigma$ confidence. This level of SZ is consistent with expectations given other SZ measurements of low mass systems and the estimated mass of the host halos of radio-active AGN. The data are well fit by an SZ effect term that is constant for all 1.4~GHz flux density bins. See Section~\ref{discussion} for further discussion of these results, including a comparison of the SZ effect measured for this sample with that measured for the \citet{best2012} sample.

The spectral index from 1.4 to 4.8~GHz is consistent with the spectral index from 1.4 to 218~GHz for sources across the full range of 1.4~GHz flux densities, as seen in Figure \ref{fig:uberspec}.  The AGN parameters $\alpha_1$, and $\alpha_0$ are constrained most at
intermediate to high $S_{1.4}$,  where
the SZ effect represents a smaller fraction of the flux density (though still $\sim$20 \textendash\ 30\%) 
than in fainter bins. The estimated
effective AGN spectral index at the highest 1.4~GHz flux density bin is \ParAlphaWithSZ, consistent with optically thin,
``steep spectrum'' synchrotron emission, as observed in other studies.  As indicated by the $\alpha_1$ parameter, the AGN
spectral index varies with 1.4~GHz flux density such that at the lowest 1.4~GHz flux density bin (S$_{1.4} \approx 6$~mJy),
$\alpha = 0.5$. This is expected for a source
population composed of some flat spectrum AGN and some steep spectrum
AGN where the relative prevalence of one population over the other varies with low frequency flux density, as is further discussed in Section \ref{flatspectrumsources}.  The $\alpha=0.55\pm0.03$ measured for the median SED of the \citet{best2012} sample in Section \ref{agnsedresults} lies in the middle of the range for this model. However, for the distribution of NVSS flux densities of the \citet{best2012} sample, the average expected $\alpha$ according to this analysis would be $0.467\pm0.005$ (statistical uncertainty from MCMC).    
The imposition of the requirement for optical counterparts on the \citet{best2012} sample may influence the population selected such that the ensemble synchrotron spectra differ from the full radio source sample.  For example,  \citet{best2012} preferentially exclude high redshift galaxies, whose counterparts are not easily identifiable in SDSS.  If there is some evolution in the spectra or the relative importance of flat-spectrum and steep-spectrum sources, then this could account for the differences we observe between the preferred spectral indices in the two analyses.  

The best fitting bolometric luminosity of the radio sources in this sample ($\log_{10}(L_{IR}/L_{\odot}) = 9.96 \pm 0.08$) is significantly higher than the best fitting bolometric luminosity for the lower redshift \citet{best2012} sample ($\log_{10}(L_{IR}/L_{\odot}) = 8.7 +0.1/-0.3$), likely indicating redshift evolution in the dust emission from the host galaxies.

\subsection{Flat and Steep Spectrum AGN and Synchrotron Spectral Steepening} \label{flatspectrumsources}

Because the sources in this work are selected at 1.4~GHz (and sources detected at millimeter waves are excluded), models describing populations of low frequency sources apply to this sample more directly than those describing sources at higher frequencies.  \citet{massardi10} model source populations from $1-5$~GHz.  As seen in previous work \citep{dezotti10}, 
they find that steep spectrum sources dominate the 1.4~GHz source counts at all flux density levels, although with increasing contributions from BL Lac sources (characterized by flat spectra) at low flux densities ($\la10$mJy) and from flat spectrum radio quasars at high flux densities ($\ga500$mJy).  Thus for this paper (which only includes sources below 200~mJy), we would expect to see a population of flat spectrum sources become more important at low 1.4~GHz flux densities, with the caveat that the spectral behavior at high frequencies may diverge from that predicted by the counts based on studies up to 5~GHz.  This population would introduce a flattening of the inferred spectral index at low 1.4~GHz flux densities. We do indeed find that the population-averaged spectral index from 1.4~GHz to millimeter frequencies is flatter at low 1.4~GHz flux densities compared to the equivalent spectral index for sources with high 1.4~GHz flux densities (see Section \ref{sec:modeling}).  Conversely, electron aging could cause a steepening of the average synchrotron spectrum, although some models predict that only the brightest blazars would tend to have such steepening at wavelengths longer than the sub-millimeter regime (\citealp{ghisellini1998}; for a list of references in support of and opposed to these models, see \citealp{dezotti10}).  

\begin{figure}
	\centering
	\includegraphics[width=84mm]{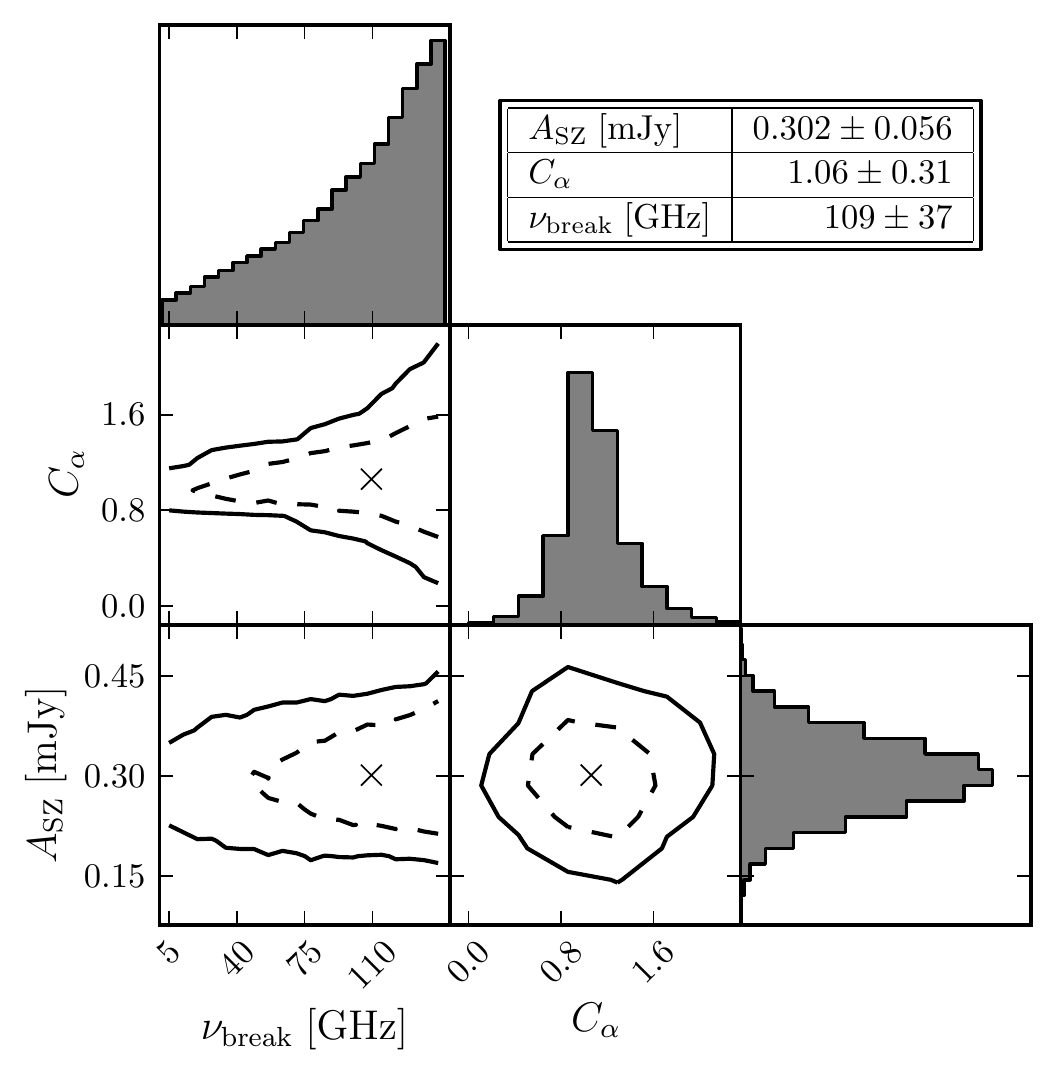}
\caption{The posterior distributions of the key model parameters for the model in which the average spectra is allowed to steepen, with best fit values listed. The histograms show the single-parameter distributions for each parameter. The dashed lines show the 68\% confidence regions, and the solid lines show the 95\% confidence regions. The parameters correspond to the amplitude of the SZ from ionized gas in AGN dark matter halos ($A_{SZ}$), the amount by which the synchrotron spectrum index steepens ($C_{\alpha}$), and the frequency at which the spectrum steepens ($\nu_{break}$).  This frequency, $\nu_{break}$, is not well constrained by our data and tends to prefer a location for the steepening above the lowest ACT band at 148~GHz.  The amplitude of the SZ effect is significantly non-zero even for models in which steepening is allowed.
 \label{steepeningparams}}
\end{figure}

In order to test for the effects of spectral steepening or of the contribution from flat spectrum sources, we introduce another parameter, C$_{\alpha}$, defined such that $\alpha(\nu>\nu_{break}) = C_{\alpha} \alpha(\nu < \nu_{break})$, where C$_{\alpha}$ can vary from $-10$ to 10 (and thus can capture a steepening, flattening, or inverted spectrum).  We also introduce a parameter for the location of the break in the synchrotron spectrum, $\nu_{break}$.  The best fit parameters for this model are shown in Figure \ref{steepeningparams}.  The location of the break in the spectrum is not well constrained by our data and tends toward the ACT frequencies. Including these new parameters does not affect the best-fit values for any of the other model parameters: $A_{SZ}$ becomes \ParASZWithSZandSteep~mJy ~(\ParASZWithSZandSteepSignif $\sigma$ significance).   When including the entire sample, from low 1.4~GHz flux densities to high, the best fit value for this new parameter is C$_{\alpha}=$ \ParSWithSZandSteep.  The $\chi^{2}$ value for the fit is \ModelChiSqWithSZandSteep~for \ModelDOFWithSZandSteep~degrees of freedom (PTE = \ModelPTEWithSZandSteep), so the fit is not significantly improved by the addition of this parameter. If we only model the sources with $S_{1.4}<10$~mJy, the value for C$_{\alpha}$ does not change significantly. The emergence of a population of flat-spectrum sources could in principle contribute to the term in our model attributed to the SZ effect at low 1.4~GHz flux densities, but this scenario is not preferred by our data when we introduce C$_{\alpha}$. If we completely replace the SZ effect parameter from the model by this C$_{\alpha}$ parameter, the resulting $\chi^2 =~$\ModelChiSqWithoutSZandSteep, with \ModelDOFWithoutSZandSteep~degrees of freedom.  Finally, we saw in Section \ref{sec:modeling} that for the model in which we include an SZ effect parameter, the average spectral index from 1.4~GHz to 4.8~GHz agrees well with the average spectral index from 1.4~GHz to 218~GHz for a given 1.4~GHz flux density bin, implying that the simpler model without spectral steepening adequately describes the data.

\subsection{Galaxy Groups and Clusters}
\label{gmbcg_sample}

Radio AGN occasionally reside in massive galaxy clusters and groups. To check whether the SZ effect inferred from the modeling of the data is dominated by galaxy clusters instead of arising from the more typical, lower-mass environments that host most AGN, we identify and remove radio sources associated with optically-selected clusters.  

The Gaussian Mixture Brightest Cluster Galaxy (GMBCG)
catalog\footnote[5]{\url{http://home.fnal.gov/~jghao/gmbcg_sdss_catalog.html}}
\citep{hao10} contains more than 55,000 optically-selected galaxy
clusters in SDSS Data Release 7 with redshift range
$0.1<z<0.55$. While the radio source redshift distribution extends to higher redshift (see Figure \ref{fig:zdist}), we can estimate the contribution of clusters to the SZ effect signal associated with radio sources based on this lower redshift sample.  \citet{thesis} found that the number of radio sources per unit cluster mass does not evolve strongly with redshift.  The number density of radio sources does evolve with redshift, implying that if anything the fraction of radio loud AGN in clusters at redshifts above $z\sim0.5$ is lower than today.  Thus a significant fraction of the contribution of clusters' SZ effect to the average SZ effect associated with radio AGN halos should be captured by investigating these low to intermediate redshift systems.  

Within the area of overlap between the ACT and FIRST surveys, there are \GMBCGInReg\ GMBCG
clusters. Comparing this subset with our 1.4~GHz sample, there were \GMBCGAssocSourcesPreThin\ sources within a $1\arcmin$ projected radius of a GMBCG cluster (out of \TotalStackedSources\ total sources in the sample).  Of these, only \GMBCGAssocSourcesPostThin\ would have been included in the stacking analysis; the others would have been excluded because they contain multiple radio sources. Excluding these sources near clusters from the analysis changes the $S_{148}$ average values by \GMBCGFracChangeAROne, which is small compared to the errors. The model parameters (Section \ref{sec:modeling}) are not significantly affected by the exclusion of sources that are near clusters, indicating that the $A_{SZ}$ term is not dominated by the richest systems. 

\section{Discussion}
\label{discussion}
\subsection{SZ effect of radio source host halos}\label{szdiscussion}

As presented in Sections \ref{agnsedresults} and \ref{sec:modeling}, there is evidence for the SZ effect from the hot atmospheres associated with the halos hosting the radio-loud AGN at the $3\sigma$ level for the sample selected from \citet{best2012} and at the \ParASZWithSZSignif$\sigma$ level for the sample selected from \citet{kimball08}.
In order to compare our results between samples and to other
measurements of the SZ effect in the literature, we must convert the detected SZ amplitude to an integrated SZ
signal describing the average SZ effect of the halos hosting the radio AGN.  This conversion requires an assumed angular shape of the SZ effect in order to account for its convolution with the ACT beam.  For a detailed discussion of our method, assumptions and estimated uncertainties in calculating the integrated SZ signal from our measurements, see Appendix \ref{scalingforprofile}.

For the \citet{best2012} sample, we calculate an integrated $E(z)^{-2/3} D_{A}(z)^{2} Y_{200}$ = \YIntBest $h_{70}^{-2}$ Mpc$^2$ where $D_A$ is the angular diameter distance.  For the \citet{kimball08} sample, we calculate an integrated $E(z)^{-2/3} D_{A}(z)^{2} Y_{200}$ = \YIntegratedAmpl $h_{70}^{-2}$ Mpc$^2$.  The systematic uncertainties quoted include uncertainties on the profile and on the profile concentration parameter, and for the \citet{kimball08} sample also include uncertainties in the redshift distribution of the radio sources.  The integrated $Y$ for the lower redshift \citet{best2012} sample thus exceeds that of the mostly higher redshift \citet{kimball08} sample, although at low ($<2\sigma$) significance.  This may indicate evolutionary growth in the typical halo mass and associated gaseous atmosphere of radio galaxies from high redshift to today. Because the the Compton parameter is integrated to $\theta_{200}$, it follows that the $Y_{200}$ thus determined depends systematically on the mass ($Y_{200} \propto \theta_{200}^2$  and $\theta_{200} \propto M_{200}^{1/3}$) due to this geometric factor.  While the uncertainty on the mass is propagated through this relation to the uncertainty quoted on the measured $Y_{200}$, this can also introduce a systematic uncertainty that would follow the same relation. Since we have assumed the same mass for both samples, if the high redshift halos have lower average mass, the underlying difference in the integrated $Y$ between the high and low redshift samples would be intrinsically larger than this observed difference.

\begin{figure}
	\centering
	\includegraphics[width=84mm]{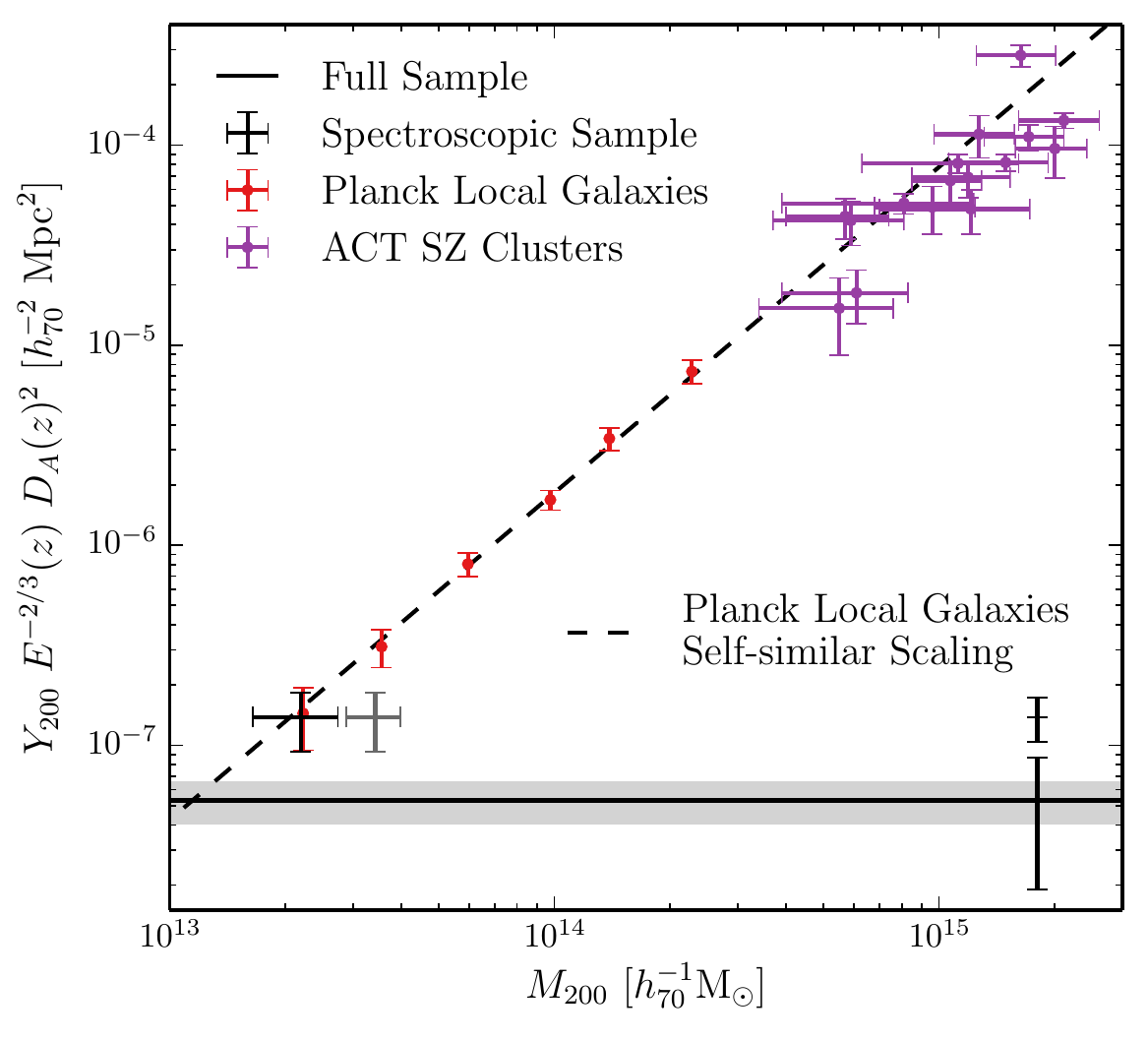}
         \caption{The SZ observable versus mass.   The black point corresponds to the estimated integrated $Y$ parameter that describes the SZ effect of halos hosting radio sources in the model that best fits our data for the sample from \citet{best2012}, and the black error bar at the right of the plot that corresponds to this integrated $Y$ value indicates the systematic uncertainty as estimated in Appendix \ref{scalingforprofile}.  The mass value of the black point corresponds to the average mass of halos hosting similarly selected radio sources as determined from their weak lensing signal in \citet{mandelbaum09}.  Because the averages are calculated over a broad distribution in $Y$ and $M$, and because $Y \propto M^{5/3}$ for self-similar scaling, we also indicate with a gray point the mass from \citet{mandelbaum09} shifted to $\langle{M^{5/3}}\rangle^{3/5}$according to expectations for the mass distributions outlined in Appendix \ref{distributionappendix}. The solid line indicates the estimated integrated $Y$ parameter that describes the SZ effect of halos hosting radio sources in the model that best fits our data for the sample from \citet{kimball08}, which has a redshift distribution that extends to much higher redshift than the sample from \citet{best2012}.  The gray region indicates the statistical uncertainty on that measurement, and the black error bar at the right of the plot that corresponds to this integrated $Y$ value indicates the systematic uncertainty as estimated in Appendix \ref{scalingforprofile}.  The purple data indicate the measurements for ACT SZ-detected galaxy clusters with dynamical mass measurements \citep{sifon}. The red data indicate results from \citet{plancklocalbrightgalaxies} from measuring the SZ effect of local bright galaxies through stacking analyses, and the dashed line indicates the scaling relation derived in that study. That the black point agrees well with the data at similar masses from \citet{plancklocalbrightgalaxies} shows consistency between our measurements.  The $Y$ measurement for the higher redshift sample is lower than for the lower redshift sample, which may indicate evolution in the typical radio source host halo. \label{fig:comparemasses}}
\end{figure}

Figure \ref{fig:comparemasses} shows the integrated {\it Y} parameters for both samples, with the low redshift \citet{best2012} integrated $Y$ parameter plotted against the mass measurement for a similarly low redshift sample from a weak lensing analysis by \citet{mandelbaum09}.  Because these measurements probe a wide range in mass and $Y$ and for self-similar scaling $Y\propto M^{5/3}$, we also plot the mean integrated {\it Y} parameter against $\langle{M^{5/3}}\rangle^{3/5}$, which we calculate in Appendix \ref{distributionappendix}.
If the true masses of these systems are as implied, this study has measured the SZ effect for some of the lowest mass halos to date, as most other work investigating the stacked SZ properties of galaxies and groups targeted $\sim10^{14}$M$_{\odot}$ systems \citep{hand, planckmaxbcg, sehgalmaxbcg}.  
\citet{plancklocalbrightgalaxies} probe down to a similar mass regime as we do and are also consistent with extrapolations from {\it Y}--M scaling relations based on high (galaxy cluster) mass halos.  While these previous studies bin according to mass proxy, this work 
computes an average value over what is likely a wide range of halo masses (see Section \ref{galaxyformation}).  There is good agreement among our SZ measurement for the low redshift radio galaxies, the weak lensing and clustering mass measurement of radio galaxies similarly selected, and the measurements from \citet{plancklocalbrightgalaxies} for the relation between $Y$ and mass for local galaxies.  The higher redshift \citet{kimball08} sample integrated $Y$ lies below the lower redshift value, possibly indicating evolutionary growth in the typical mass of radio galaxy hosts. 

\subsubsection{Implications for galaxy formation models}\label{galaxyformation}

Recent observations and theory support an overall picture where most radio AGN are powered by radio-mode (or hydrostatic mode) accretion: hot gas from a hydrostatically supported halo ultimately cools and fuels the AGN, which in turn heats that gaseous halo, establishing a feedback mechanism.  Our measurement of the mean SZ effect integrated $Y$ parameter associated with radio-AGN-hosting halos provides provides a direct measurement of the hot gas associated with this mechanism.  

Previous studies have found that the halos that host radio-loud AGN are indeed massive, as would be implied by this feedback scenario. For example, \citet{mandelbaum09} measure two-point correlation functions and galaxy-galaxy lensing shear signals to study the dark matter halo masses of radio AGN selected from FIRST and NVSS and cross-matched with SDSS.  They find that the halos hosting radio AGN are on average twice more massive than those hosting galaxies with equivalent stellar masses but without radio activity.  Additionally, AGN without radio activity reside in galaxies that, as a population, have a very different distribution of stellar mass \citep{best2005}: radio AGN reside only in galaxies above a stellar mass threshold, but optical AGN reside in galaxies both above and below that threshold. The mean mass for halos hosting radio-loud AGN is $(2.3\pm0.6)\times10^{13} h_{70}^{-1}$ M$_{\odot}$ (for comparison, the mean mass for halos hosting optical AGN is $(1.1\pm0.2)\times10^{12} h_{70}^{-1}$ M$_{\odot}$).  
Galaxy cluster BCGs are known to be preferentially radio-loud \citep{best2007, linandmohr}, so \citet{mandelbaum09} also repeat their analysis after excluding radio AGN that lie within known optically selected clusters.  While this reduces the difference between the halo masses of radio AGN and the control sample by $\sim15\%$, they still see evidence that the dark matter halos hosting radio AGN are more massive.  

Quantifying the extent to which our measurement supports the radio mode feedback paradigm requires a prediction for the average $Y$ parameter based on the expected mass distribution for radio-loud AGN hosts.  We estimate this value to be $E(z)^{-2/3} D_{A}(z)^{2} Y_{200} \sim 3\times10^{-7}h_{70}^{-2}$ Mpc$^2$, with details of the calculation and assumptions required given in Appendix \ref{distributionappendix}. The corresponding average $Y$ for optical AGN is $E(z)^{-2/3} D_{A}(z)^{2} Y_{200} \sim 2\times10^{-8}h_{70}^{-2}$ Mpc$^2$, and for all galaxies is $3\times10^{-9}h_{70}^{-2}$ Mpc$^2$.  If radio emission from jets were present in all active galaxies such that they would be selected in this sample, the average halo mass of the population would correspond to an undetectable SZ effect signal given the current sensitivity and sample statistics. The fact that the measured $Y$ is consistent with the higher mass distribution expected for radio loud AGN provides a key consistency check of the picture that radio jets are actively providing feedback only to the most massive halos.  Even more directly, the SZ effect provides evidence for the presence of ionized gas in the halos that host radio galaxies.
We hope that the availability of this new observation will encourage the direct calculation of $Y$ from simulations, thereby providing a new constraint for testing cosmological structure formation models with radio-mode AGN feedback.

\subsection{Contribution to CMB power spectrum}
Recent large millimeter surveys have enabled more sophisticated source population and spectral index modeling of high frequency radio sources than has previously been possible \citep[e.g., ][]{tucci}.  In a practical cosmological context, understanding the millimeter wave spectral behavior of radio sources is useful for modeling and removing the contribution of radio sources to high resolution measurements of the cosmic microwave background.  Current experiments like Planck, SPT and ACT can use radio sources detected in their surveys to constrain the spectral indices of bright radio sources \citep{vieira10, marriage11, planck_ercsc, mocanu2013}.  However, detected sources may also be easily masked from cosmological analyses, while faint sources that fall below the detection thresholds of these experiments still contribute to the observed power over the angular scales they probe. A better understanding of the spectral behavior of such radio sources enables better modeling of their contribution to the power spectra and cross-power spectra for multiple frequencies \citep[e.g.,][]{mason2009}.  Characterizing the millimeter behavior of radio sources can also be useful for SZ surveys \citep{linandmohr,lin2009,bolocam,diegopartridge}, as radio sources in clusters can fill in the SZ decrement at frequencies below 220~GHz. 

We estimate the total power that these sources contribute to the CMB power spectrum at 148~GHz as
\begin{equation}
C_{\ell}^{PS} = \sum_{i=1}^n N_i (S_i - A_{SZ})^2 
\label{totalrmspower}
\end{equation}
where $N_i$ is the angular number density of sources in a 1.4~GHz flux density bin, $S_i$ is the 148~GHz flux density from the radio source emission for that bin, and $A_{SZ}$ is the flux density associated with the SZ effect. The observed average flux density at 148~GHz for each bin is $S_{i,obs} =  S_i - A_{SZ}$. We approximate the uncertainty for each flux density bin as $\sigma_{C^{PS}_i} \approx N_i S_{i,obs}^2 \sqrt{ 1/N_i + (2 \sigma_{S,i,obs} / S_{i,obs})^2 }$, and the total uncertainty on $C_{\ell}^{PS}$ is calculated as the sum in quadrature of $\sigma_{C^{PS}_i}$ over all flux density bins. The values for the total power observed (calculated using the averages for each flux density bin listed in Table \ref{first_results}) are $\ell(\ell+1)C_{\ell}^{PS}/(2\pi) = 0.14\pm0.02~\micro$K$^2$, $0.21\pm0.025~\micro$K$^2$ and $0.6\pm0.1~\micro$K$^2$ at $\ell = 3000$ for 148, 218, and 277~GHz, respectively. 

As seen Equation \ref{totalrmspower}, the power contributed by the sources at 148~GHz is partially suppressed by the SZ effect term, and similarly the 277~GHz power is partially enhanced by the SZ effect.   The total power can be broken into components as 
\begin{equation}
C_{\ell}^{PS} 
= C_{\ell}^{\mathrm{A \times A}} + C_{\ell}^{\mathrm{SZ \times SZ}} + C_{\ell}^{\mathrm{A \times SZ}}
\end{equation}
where $C_{\ell}^{\mathrm{A \times A}} = \sum_{i=1}^n N_i S_i^2$ accounts for the power contributed by the radio sources' emission, $C_{\ell}^{\mathrm{SZ \times SZ}} \approx  N_{tot} A_{SZ}^2$ accounts for the power contributed by the SZ effect of the hosts of the radio sources, and $C_{\ell}^{\mathrm{A \times SZ}} = - 2 \sum_{i=1}^n N_i S_i A_{SZ,i}$ is the cross-power.  In order to calculate the $C_{\ell}^{A \times A}$ term, we add to the average 148~GHz flux density for each bin a constant corresponding to the SZ effect contribution at 148~GHz, $A_{SZ}$, for the best fit model ($A_{SZ}=$\ParASZWithSZ~mJy; see Section \ref{sec:modeling}).  Thus the contribution of radio sources, corrected for the SZ effect of their hosts, to the CMB power spectrum at 148~GHz becomes $\ell(\ell+1)C_{\ell}^{\mathrm{A \times A}}/(2\pi)=0.37 \pm0.03~\micro$K$^2$ at $\ell=3000$.  This contribution is in general agreement with that predicted by models: for sources with  $0.3$~mJy~$ <S_{148}<2$~mJy, \citet{tucci} model predictions range from 0.29 to 0.32~$\micro$K$^2$ and \citet{dezotti05} predict 0.35~$\micro$K$^2$.  
Similarly, the contribution to the power spectrum from radio sources at 277~GHz after correcting for the SZ effect (including both its frequency dependence given in Section \ref{sedmodel} and the effect of the filter discussed in Appendix \ref{scalingforprofile}) is $\ell(\ell+1)C_{\ell}^{\mathrm{A \times A}}/(2\pi)=0.25\pm0.07~\micro$K$^2$ at $\ell=3000$.  The contribution from radio sources to the 218~GHz power spectrum is unaffected by the SZ effect and is estimated to be $\ell(\ell+1)C_{\ell}^{\mathrm{A \times A}}/(2\pi) = 0.21\pm0.025~\micro$K$^2$ at $\ell=3000$. 
For comparison, the prior on the contribution of radio sources to the CMB power spectrum measured by ACT at 148~GHz that was used by \citet{dunkley2013} 
for sources $S_{148}<15$~mJy  
was 2.9$\pm0.4~\micro$K$^2$ at $\ell=3000$.  The estimated power of sources in the SPT power spectrum after masking to a level of 6~mJy is $1.3\pm0.2~\micro$K$^2$ at $\ell=3000$ \citep{reichardt2012}.  The amount of power that we estimate the radio sources in this study ($5.0$~mJy~$<S_{1.4}<200$~mJy, $0.3$~mJy~$ <S_{148}<2$~mJy) contribute after correcting for the SZ effect is $0.37\pm0.04~\micro$K$^2$, which accounts for a fraction of the assumed priors ($\sim25\%$ for SPT and $\sim10\%$ for ACT).  

We can also evaluate the power contributed by the SZ effect of the halos that host radio sources and the cross-power between the SZ effect and AGN emission components. Because we bin in terms of radio source power and not SZ effect amplitude, there is the assumption that $\langle A_{SZ}^2 \rangle \sim \langle A_{SZ} \rangle^2$.  The SZ effect term becomes $\ell(\ell+1)C_{\ell}^{\mathrm{SZ \times SZ}}/(2\pi) = 0.06\pm0.004~\micro$K$^2$ (statistical uncertainty) at $\ell=3000$.  For comparison, \citet{sievers2013} find that the contribution of the thermal SZ effect to the ACT CMB power spectrum at $\ell=3000$ is $3.4\pm1.4~\micro$K$^2$.  The cross-power term is $\ell(\ell+1)C_{\ell}^{\mathrm{A \times SZ}}/(2\pi) = -0.29 \pm 0.07~\micro$K$^2$.

\section{Conclusions}
\label{conclusions}

We have investigated the ensemble millimeter properties of 1.4~GHz selected radio sources by stacking 148, 218, 277, 600, 857 and 1200~GHz data from ACT and {\it Herschel} on the positions of radio sources selected from two joint catalogs of FIRST and NVSS 1.4~GHz sources.  Although most of the radio sources are AGN, whose spectra are expected to fall with increasing frequency at millimeter wavelengths, we see evidence for an average rising spectrum. This observed spectral inversion in the ACT bands is attributed to the SZ effect of the halos that host the AGN.  We construct SEDs for a sample of radio sources \citep{best2012} that have optical counterparts and thus known redshifts and classifications as either AGN or SFGs.  In order to constrain the contribution to the SED from dust emission, we also stack data from {\it Herschel} surveys, and to better constrain the synchrotron spectrum we include 5~GHz data from PMN and a GBT survey. The AGN SED is well fit by a model that includes synchrotron emission, the SZ effect and dust emission.  The SFG SED is well fit by a model that includes synchrotron emission, free-free emission, CO line emission and dust emission. 

Using a larger catalog enables the construction of SEDs for sources binned in 1.4~GHz flux density. The radio source sample from \citet{kimball08} was constructed without optical counterparts, and while spectroscopic redshifts are lacking, the sample includes greater numbers of sources (\TotalStackedSources\ used in this study) and extends to higher redshifts.  Even though we exclude bright ACT sources, the binned 1.4~GHz selected radio sources are detected in the ACT data.  The ensemble averaged millimeter synchrotron spectral index ($S\propto \nu^{-\alpha}$) for AGN is \ParAlphaWithSZ\ for the bright ($\sim100$~mJy) sources, with evidence for flattening toward fainter 1.4~GHz source flux densities.   The best-fit model includes an SZ contribution detected at the \ParASZWithSZSignif$\sigma$ level. 

The detection of the SZ effect in the radio galaxies' millimeter wavelength SED provides evidence that, on average, galaxies that host radio AGN support hot gaseous halos.  This is important in the context of the growing support for the overall picture that most radio AGN are powered by radio-mode accretion, where AGN are fueled by the accretion of gas ultimately from the hot hydrostatically supported halo.  In this picture, the radio jets from the AGN in turn provide energy that counters the cooling of the gas, shutting off the AGN fuel supply by establishing a feedback loop.   

We have also calculated the contribution of these radio sources to the CMB power spectrum.  At 148~GHz, after correcting for the SZ effect, the sources contribute $\ell(\ell+1)C_l^{\mathrm{A \times A}}/(2\pi)=0.37 \pm0.03~\micro$K$^2$ at $\ell=3000$, where the uncertainty is dominated by the uncertainty on the contribution to the SZ effect to the measured flux densities.  The cross-power between the SZ effect and the radio source emission at 148~GHz is $\ell(\ell+1)C_{\ell}^{\mathrm{A \times SZ}}/(2\pi) = -0.29 \pm 0.07~\micro$K$^2$.

\section*{Acknowledgements}
This work was supported by the U.S. National Science Foundation through awards AST-0408698 and AST-0965625 for the ACT project, as well as awards PHY-0855887 and PHY-1214379. Funding was also provided by Princeton University, the University of Pennsylvania, and a Canada Foundation for Innovation (CFI) award to UBC. ACT operates in the Parque Astron\'omico Atacama in northern Chile under the auspices of the Comisi\'on Nacional de Investigaci\'on Cient\'ifica y Tecnol\'ogica de Chile (CONICYT). Computations were performed on the GPC supercomputer at the SciNet HPC Consortium. SciNet is funded by the CFI under the auspices of Compute Canada, the Government of Ontario, the Ontario Research Fund -- Research Excellence; and the University of Toronto. M.G. and T.M. acknowledge support from Johns Hopkins University.  R.D., P.A., F.R. and G.M. received funding from the Chilean grants FONDECYT 11100147, and BASAL (CATA).  This research made use of Astropy, a community-developed core Python package for Astronomy \citep{astropy}.

\bibliography{mn-jour,paper_master_arxiv}
\bibliographystyle{mn2e_hacked}

\appendix
\section{Calculation of integrated $Y$ parameter from measured parameter $A_{SZ}$}
\label{scalingforprofile}

Our measured deviation in the ACT 148~GHz flux densities of radio AGN
due to the SZ effect $A_{SZ}$ is described by the equation
\begin{equation}
A_{SZ}  = 2\pi I_0  \mid g(148~{\rm GHz}) \mid   \int  \Phi(\theta) \hat{y}_{gas}(\theta) d\theta,
\label{eqn:deltaS}
\end{equation}
where $I_0$ is the CMB intensity, $g$ is the SZ frequency dependence
(Equation \ref{szspec}), and $\Phi(\theta)$ is the matched filter, all
corresponding to the 148~GHz band. Azimuthal symmetry of the projected SZ effect has been assumed. 
The function $y_{gas}(\theta)$ is the dimensionless Compton parameter. The
hat indicates it has been modified to account for the effects of the
ACT beam and pixelization. This profile is related to the gas
pressure profile $P(r)$ through the line of sight projection according to
\begin{equation}
y_{gas}(\theta) = \frac{\sigma_T}{m_e c^2} \int ds P\left(\sqrt{s^2 + (R_{500}\theta/\theta_{500})^2}\right),
\label{eqn:comptony}
\end{equation}
where $\sigma_T$ is the Thomson cross section, $m_e c^2$ is the rest
mass energy of the electron, and $R_{500}$ and $\theta_{500}$ refer to
the radius and angle, respectively, within which the average density
of the dark matter halo is 500 times the critical density,
$\rho_{crit}(z)=3H^2(z)/8\pi G$. Though we report quantities in this
work referenced to the larger radius $R_{200}$, characteristic of the
halo's virial radius, we adopt the $R_{500}$ scaling for $P(r)$
because we have chosen to model the pressure according to the
``universal pressure profile'' (UPP) of
\cite{arnaud10}. The UPP is derived from X-ray observations
of clusters, for which $R_{500}$ is a more natural scale.  \citet{sun2011} show that an average pressure profile for galaxy groups agrees well with the UPP.  This profile is also used by \citet{plancklocalbrightgalaxies} to model the SZ effect of the halos hosting local bright galaxies.

To estimate a characteristic $R_{500}$ 
for the radio AGN, we use the lensing mass estimate from
\cite{mandelbaum09} ($M_{200}= 2.3 \pm 0.6 \times 10^{13}
h_{70}^{-1}$M$_\odot$). 
For $R_{500}$ and other relevant quantities (e.g., angular diameter distance), we calculate medians and means based
on the redshift distributions shown in Figure \ref{fig:zdist}.
To use the mass
estimate from \cite{mandelbaum09} with the UPP $R_{500}$-scaled
profile, we convert $M_{200}$ to $M_{500} \approx 1.5 \times 10^{13}
h_{70}^{-1}M_\odot$ assuming the dark matter follows a
\cite{nfw1997} (NFW) profile, using the concentration$-$mass relation from \cite{duffy08}. Estimated in this way,
$R_{500}= \RfhBest h_{70}^{-1}{\rm Mpc}$ ($\theta_{500} = \TfhBest'$) for the
sample of \cite{best2012}, and $R_{500} = \Rfh
h_{70}^{-1}{\rm Mpc}$ ($\theta_{500}= \Tfh'$) for the higher
redshift sample of \cite{kimball08}. The redshift
distribution of halos for which \cite{mandelbaum09} estimated a
mean $M_{200}$ is best matched to the lower-redshift sample of
\cite{best2012}. Lacking a mass estimate for radio
AGN-hosting halos with the higher redshift distribution of the
\cite{kimball08} sample, we have simply used the same
$M_{200}$ to estimate $R_{500}$ for both samples. 

\begin{figure}
	\centering
	\includegraphics[width=84mm]{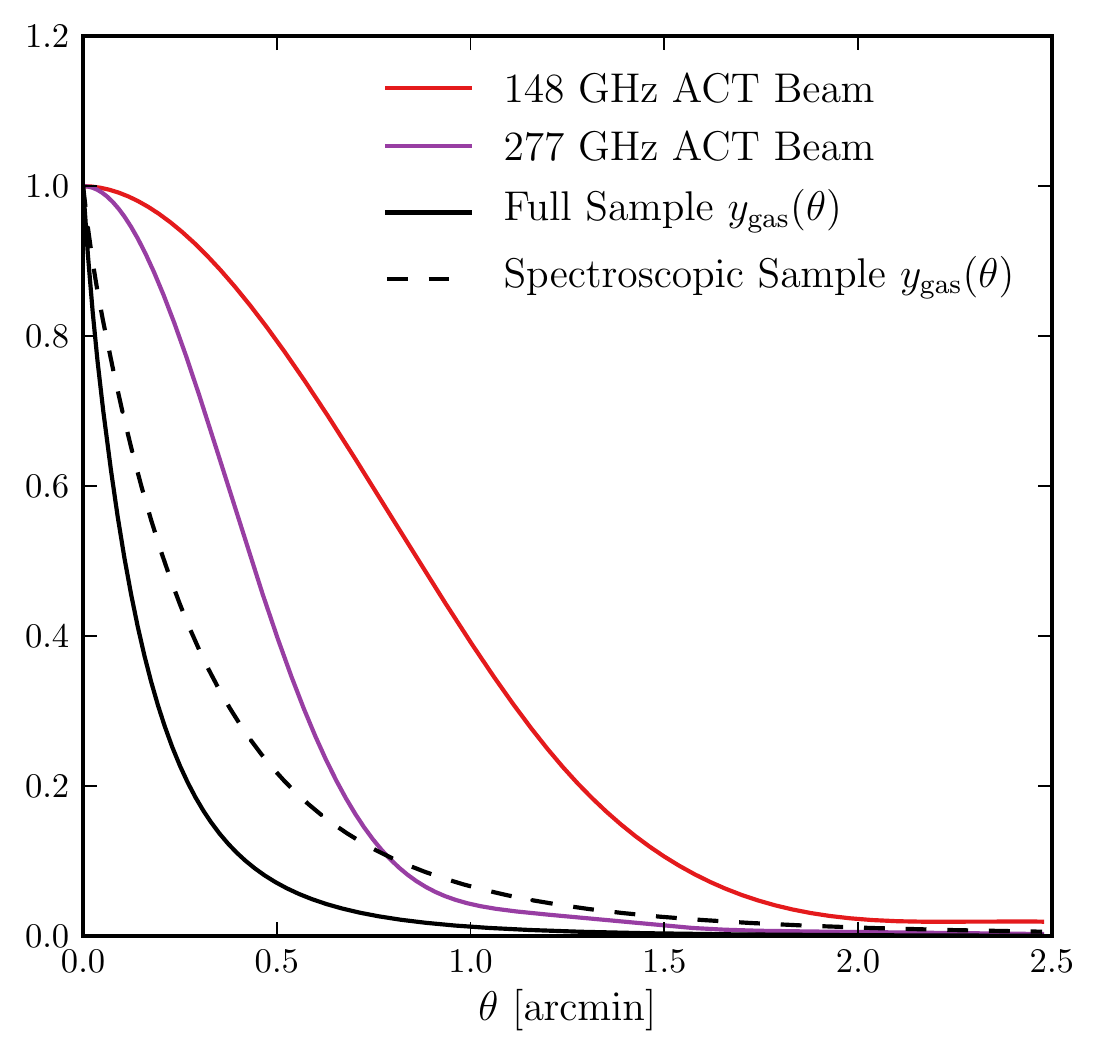}
\caption{The angular profiles of the projected SZ effect pressure and for the ACT beams.  The dashed line shows the profile assumed for the sample from \citet{best2012} (median $z=0.30$) and the solid line shows the profile assumed for the sample from \citet{kimball08} (median $z=1.06$).  The red line shows the ACT 148~GHz beam profile, and the purple line shows the ACT 277~GHz beam profile. \label{fig:beams}}
\end{figure}

Unfiltered brightness profiles proportional to $y_{gas}(\theta)$ together
with the ACT 148~GHz and 277~GHz beams are shown in Figure \ref{fig:beams}. Since
the ACT data were filtered to optimally recover point source flux, we
must correct for the bias arising from the interaction of the extended
SZ intensity profile with the matched filter. In general, the measured
$A_{SZ}$ will be lower than the total ``SZ flux density'' of the
extended profile.  To estimate the bias correction, we apply the same
filter kernels to maps containing beam convolved, simulated pressure
profiles, input with known amplitude, and examine the resulting output
amplitudes. Using simulated profiles placed according to the redshift distributions assumed for each sample, we can construct the estimated distribution of
multiplicative factor $f$ required to correct for this bias. This bias factor allows us to write $A_{SZ}$, as in Equation \ref{eqn:deltaS}, in terms of the integrated Compton $Y$ parameter, $Y_{200}$, defined as the Compton $y_{gas}(\theta)$ integrated to $\theta_{200}$:
\begin{equation}
A_{SZ} \approx  I_0 \mid g(148~{\rm GHz}) \mid f^{-1}  Y_{200}
\end{equation}
We find that for the lower-redshift \cite{best2012} sample, the median
$f$ corresponds to factors of \AROneMedCorrFacBest\ and
\ARThreeMedCorrFacBest\ for the 148~GHz and 277~GHz bands
respectively. For the full sample of \cite{kimball08} the mean 
corresponding factors are \AROneMeanCorrFac\ and \ARThreeMeanCorrFac. The
ratio of the 148~GHz to 277~GHz bias correction factors were used in
Sections 3 (with a value of \AROneToARThreeMedCorrBest) and 4 (with a value of \AROneToARThreeMeanCorr) to model the median and mean SZ effect,
respectively. In the remainder of this section, we use the bias
correction at 148~GHz to relate the estimates of $A_{SZ}$ for the two
AGN samples to corresponding estimates of $Y_{200}$.
For the \citet{kimball08} sample, these assumptions yield an integrated $E(z)^{-2/3} D_{A}(z)^{2} Y_{200}$ = \YIntegratedAmplnosys $h_{70}^{-2}$ Mpc$^2$ where $D_A$ is the angular diameter distance and the uncertainties quoted are statistical only.  For the \citet{best2012} sample, these assumptions, along with the redshifts measured for the sources, yield an integrated $E(z)^{-2/3} D_{A}(z)^{2} Y_{200}$ = \YIntBestnosys $h_{70}^{-2}$ Mpc$^2$, where the uncertainties quoted are statistical only.

In order to characterize the level of additional systematic uncertainties, we varied the assumptions made in this calculation about the radio source redshift distribution, the typical pressure profile and the typical halo concentration. To estimate the $1\sigma$ systematic uncertainty contributed by each of these assumptions, we calculate the difference between the values given the alternative models described below and the values calculated using our fiducial model.  To calculate a total systematic uncertainty for each sample, we add these systematic uncertainties in quadrature.
  If we adopt the redshift distribution of the radio sources given by \citet{ho2011}, the value for the \citet{kimball08} sample becomes $E(z)^{-2/3} D_{A}(z)^{2} Y_{200} =~$\YIntegratedAmplHo\ $h_{70}^{-2}$ Mpc$^2$, which is in agreement with the value calculated above for the redshift distribution as reported by \citet{dezotti10}.  Adopting the halo concentration--mass relation from \citet{neto2007} yields $E(z)^{-2/3} D_{A}(z)^{2} Y_{200} =$ \YIntBestNeto\  $h_{70}^{-2}$ Mpc$^2$ for the \citet{best2012} sample and $E(z)^{-2/3} D_{A}(z)^{2} Y_{200} =$ \YIntegratedAmplNeto\ $h_{70}^{-2}$ Mpc$^2$ for the \citet{kimball08} sample. 
 Finally, if we adopt the best-fit profile of
\cite{planckpressureprofile}, who measured
the SZ pressure profiles of 62 massive, low redshift clusters, we find
a resulting value of $E(z)^{-2/3} D_{A}(z)^{2} Y_{200}$ = \YIntBestPlanck\ $h_{70}^{-2}$ Mpc$^2$ for the \citet{best2012} sample and $E(z)^{-2/3} D_{A}(z)^{2} Y_{200}$ = \YIntegratedAmplPlanck\ $h_{70}^{-2}$ Mpc$^2$ for the \citet{kimball08} sample. This pressure
profile parameterization is more extended than the fiducial \citet{arnaud10}
profile, so a given estimate of the amplitude of the signal yields a larger value for the integrated $Y$ parameter.  
 
 In summary, we find an integrated $E(z)^{-2/3} D_{A}(z)^{2} Y_{200}$ = \YIntegratedAmpl $h_{70}^{-2}$ Mpc$^2$ for the \citet{kimball08} sample and an integrated $E(z)^{-2/3} D_{A}(z)^{2} Y_{200}$ = \YIntBest $h_{70}^{-2}$ Mpc$^2$ for the \citet{best2012} sample.

\section{Estimated halo mass distribution for radio loud AGN hosts} \label{distributionappendix}
We have measured the mean integrated Compton parameter $Y$  for hosts of radio-loud AGN. From self-similar arguments we expect this parameter to be related to the host's halo mass $M_{200}$ as  $Y_{200} E^{-2/3}(z) D_A^2(z)  \propto M_{200}^{5/3}$ \citep{kaiser1986}.  In this appendix we estimate the mean $M_{200}$ and $M_{200}^{5/3}$ for local ($z<0.6$) radio-loud AGN hosts, corresponding to the redshift range of the spectroscopic sample \citep{best2012} used in Section \ref{sedanalysis}. Schematically this is done by using the stellar mass function for all galaxies $\phi(M_\star)$ \citep{cole2001} scaled by the radio-loud AGN fraction $f_{rad}(M_\star)$ \citep{best2005} and then using an approximate functional form for the stellar-to-halo mass relation $M_{200}(M_\star)$ from lensing \citep{mandelbaum09} to compute the mean $M_{200}$ and $M_{200}^{5/3}$. A similar treatment is carried out for optically-selected AGN. This allows us to  approximate what mean Compton $Y$ is expected for AGN hosts. Perhaps more importantly, it allows us to evaluate the level of correction needed in comparing the mean $Y_{200} E^{-2/3}(z) D_A^2(z)$, taken over the entire AGN host population (from $\langle M_{200}^{5/3} \rangle$), to the mean halo mass $\langle M_{200} \rangle$ from lensing or clustering studies, as in Figure \ref{fig:comparemasses}. 

With a prescription for the Initial Mass Function (IMF) and stellar synthesis, one can calculate the stellar mass function from galaxy surveys \citep[e.g.,][]{cole2001,bell2003, baldry2012, moustakas2013}. We use  the simple Schechter function from \cite{cole2001} for the form of the local mass function (Figure 18, Table 4).The form of Schechter function $\phi(M_\star)$, which gives the density of galaxies per stellar mass ($h^3$/Mpc$^3 \times h^2/$M$_\odot$), has the form:
\begin{equation}
\phi(M_\star) dM_\star = \phi^*_\star \left(\frac{M_\star}{M^*_\star}\right)^\alpha \exp(-M_\star/M^*_\star) \frac{dM_\star}{M^*_\star}
\label{eqn:stellarmassfunc}
\end{equation}
where $M^*_\star = 3.43 \times 10^{10} h^{-2} {\mathrm M}_\odot$, $\alpha=-1.19$, and $\phi^*_\star=0.01 h^3$~Mpc$^{-3}$. (Recall that $H_0=100h$~km/s/Mpc, and we work in units of  $h=0.7$). These values for the Schechter function parameters are based on a fit to data assuming a Kennicutt IMF \citep{kennicutt1983}. The \cite{cole2001} Salpeter IMF -based stellar mass function results in a mean stellar masses approximately twice that associated with the Kennicutt IMF due to the higher mass tail of the Salpeter IMF \citep{salpeter1955}. More recent estimates of the stellar mass function using a Chabrier IMF \citep{chabrier2003, baldry2012,moustakas2013} appear to have a high mass tail somewhere between that of the \cite{cole2001} Kennicutt and Salpeter IMF-based stellar mass functions. However, uncertainties associated with the IMF and other relations are significant, and the Kennicutt IMF-based stellar mass function yields the best match to the mean mass measured for the radio loud AGN by the weak lensing analysis in \citet{mandelbaum09}. Evolution of the IMF, especially for the higher mass end of the distribution is minor for $z<0.6$ \citep{moustakas2013}.

\begin{figure}
\begin{center}
\includegraphics[width=84mm]{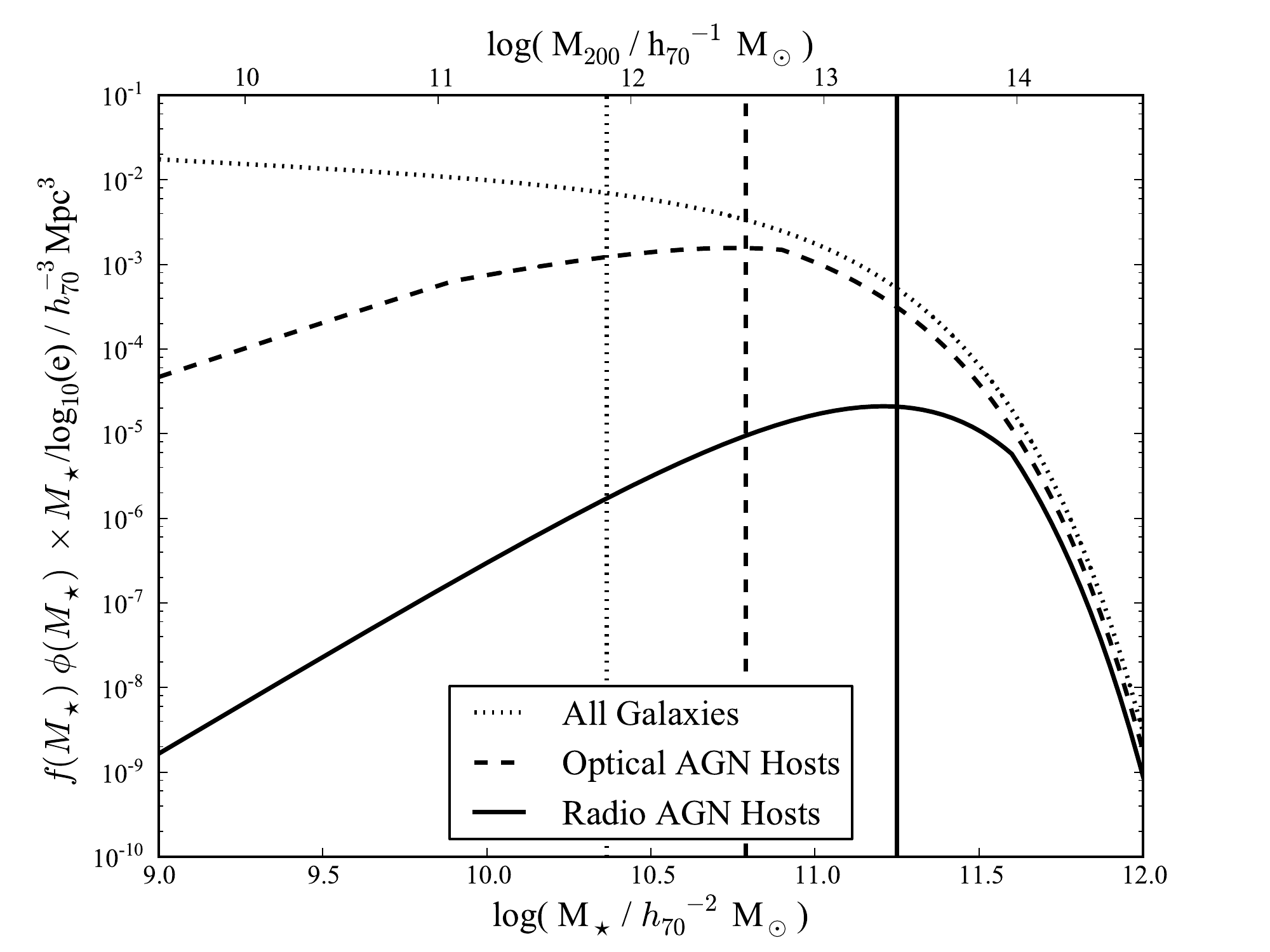}
\caption{Stellar and halo mass distribution functions (per decade in mass) for all ``control'' galaxies, galaxies with radio AGN and galaxies with optical AGN.  Vertical lines indicate the mean mass of each distribution.}
\label{fig:halodist}
\end{center}
\end{figure}

 \cite{best2005} estimate the fraction of all galaxies that host optical and radio-loud AGN as a function of the host's stellar mass.  For radio AGN ($L_{NVSS}>10^{23}$~W/Hz) the fraction scales as a power law $M_\star^{5/2}$ with an approximate final fraction of 0.3 at $\log(M_\star/$M$_\odot) \ge 11.6$ : 
\begin{equation}
f_{rad}(M_\star) = \left\{
\begin{array}{ll}
0.3 & \log(M_\star/{\mathrm M}_\odot) > 11.6\\
0.3 \times (M_\star/10^{11.6} {\mathrm M}_\odot)^{2.5} & \log(M_\star/{\mathrm M}_\odot) \leq 11.6 
 \end{array}\right.
\label{eqn:radiofrac}
\end{equation}
For the optical AGN ($L_{[{\mathrm OIII}]}>10^{5.5}$L$_{\odot}$), this function \citep[][; Figure 2, bottom-left panel]{best2005} is approximated by a twice-broken power law  :
\begin{equation}
f_{opt}(M_\star) = \left\{
\begin{array}{ll}
0.6& \log(M_\star/{\mathrm M}_\odot) > 10.9\\
0.6 \times (M_\star/10^{10.9}{\mathrm M}_\odot) & 9.9 < \log(M_\star/{\mathrm M}_\odot) \le 10.9\\
0.06 \times (M_\star/10^{9.9}{\mathrm M}_\odot)^{1.5} & \log(M_\star/{\mathrm M}_\odot) \le 9.9
\end{array}\right.
\label{eqn:opticalfrac}
\end{equation}
These fractions, $f_{rad}$ and $f_{opt}$, multiply the stellar mass function for all galaxies (Equation \ref{eqn:stellarmassfunc}) to approximate the stellar mass function for radio and optical AGN hosts, respectively. These functions are plotted in Figure \ref{fig:halodist}. Note the functions are plotted as densities per decade of stellar mass and therefore are multiplied by a factor of $M_\star/\log_{10}(e)$.

We use an approximate form of the stellar-to-halo mass relation for AGN taken from the lensing study of \cite{mandelbaum09}:
\begin{equation}
\log(M_{200})= 1.7(\log(M_\star)-10.3) + 11.5 -1.7\log(h).
\label{eqn:massconv}
\end{equation}
This form also roughly approximates mass relations from more recent work on the more general stellar-to-halo mass relation (unrelated to AGN) \cite[e.g.,][]{leauthaud2012}. With Equations \ref{eqn:stellarmassfunc}, \ref{eqn:radiofrac}, \ref{eqn:opticalfrac}, and \ref{eqn:massconv}, we compute the mean halo mass as
\begin{equation}
\langle{M_{200}}\rangle = \frac{\int M_{200}(M_\star) f(M_\star) \phi(M_\star) dM_\star}{\int f(M_\star) \phi(M_\star) dM_\star}
\end{equation}
where $f$ is either the radio or optical fraction or unity for all galaxies. The integral is taken from $M_\star=10^9$~M$_\odot h^{-2}$ to $10^{12}$~M$_\odot h^{-2}$. The mean of $M_{200}^{5/3}$, corresponding to the mass scaling of $Y E^{-2/3}(z) D_A^2(z)$, is similarly computed. The mean halo masses for all galaxies, optical AGN hosts, and radio-loud AGN hosts computed in this way are 8$\times10^{11}$~$h_{70}^{-1}$~M$_\odot$, 4$\times10^{12}$~$h_{70}^{-1}$~M$_\odot$, and 2$\times10^{13}$~$h_{70}^{-1}$~M$_\odot$, respectively.  These results are in order-of-magnitude agreement with the lensing results from \cite{mandelbaum09} in which the optical AGN host and radio-loud AGN host mean masses were $1.1\times10^{12}$~$h_{70}^{-1}$~M$_\odot$ and $2.3\times10^{13}$~$h_{70}^{-1}$~M$_\odot$, respectively.  For radio AGN, the ratio of $\langle{M_{200}^{5/3}}\rangle^{3/5}$ to $\langle{M_{200}}\rangle$ is 1.4.
  We compute the mean integrated $Y$ value using a scaling relation based on \citet{plancklocalbrightgalaxies}: $\langle{E(z)^{-2/3} D_{A}(z)^{2} Y_{200}}\rangle = 9.19\times 10^{-30} \times \langle{ M^{5/3}}\rangle $. For all galaxies, optical AGN hosts, and radio-loud AGN hosts the mean $E(z)^{-2/3} D_{A}(z)^{2} Y_{200}$ is $3\times10^{-9}h_{70}^{-2}$ Mpc$^2$, $2\times10^{-8}h_{70}^{-2}$ Mpc$^2$, and $3\times10^{-7}h_{70}^{-2}$ Mpc$^2$, respectively.
 Note that this implies that the Compton Y signal from optical AGN hosts should be $\sim100\times$ diminished in comparison with that from radio-loud AGN hosts, for the case in which the gas follows self-similar scaling with mass down to $\sim10^{12}$~$h_{70}^{-1}$~M$_\odot$.  If lower mass halos preferentially lose gas relative to this scaling, the Compton Y signal from optical AGN hosts could be even lower relative to that of radio AGN hosts. 

\end{document}